\documentclass[aps,prl,10pt,twocolumn,superscriptaddress,longbibliography]{revtex4-2}
\usepackage[utf8]{inputenc}
\usepackage{amsmath,amssymb}
\usepackage{graphicx}
\usepackage{xcolor}
\usepackage[colorlinks=true,linkcolor=blue!60!black,citecolor=blue!60!black,urlcolor=blue!60!black]{hyperref}

\newcommand{\ket}[1]{|#1\rangle}
\newcommand{\avg}[1]{\langle #1\rangle}

\begin{document}

\title{Emergent Quantum Synchronization Without Symmetry Breaking}

\author{Juan David \'Alvarez-Cuartas}
\email{juan.david.alvarez@correounivalle.edu.co}
\affiliation{Departamento de F\'isica and Centre for Bioinformatics and Photonics---CIBioFi, Universidad del Valle, 760032 Cali, Colombia}
\author{Joakim Bergli}
\affiliation{Department of Physics, University of Oslo, N-0316 Oslo, Norway}
\author{John H. Reina}
\email{john.reina@correounivalle.edu.co}
\affiliation{Departamento de F\'isica and Centre for Bioinformatics and Photonics---CIBioFi, Universidad del Valle, 760032 Cali, Colombia}

\date{\today}

\begin{abstract}
Parametric amplification of a shared cavity mode turns the light--matter coupling phase of a two-ensemble Dicke model into a control parameter. We prove analytically that a cavity-set gain band opens a window of this phase with no static superradiant instability, leaving only a Hopf bifurcation, for arbitrary couplings, spin frequencies, damping rates and ensemble sizes. A $\mathbb{Z}_2$ parity keeps the quantum steady state unique and time independent: every order parameter vanishes and the oscillation lives in a slow parity-odd Liouvillian mode. Truncated-Wigner simulations, validated against exact ones, show that above threshold its quality factor grows linearly with system size and the ensembles phase lock; below threshold, both saturate. Quantum synchronization emerges without symmetry breaking and without requiring entanglement.
\end{abstract}
\maketitle

\textit{Introduction.}---The Dicke model~\cite{Hepp1973,Wang1973} is foundational to quantum optics~\cite{Garraway2011,Emary2003,Larson2021,Alvarez2026} and many-body physics~\cite{Ritsch2013,Mivehvar2021}, exhibiting a superradiant (SR) phase transition~\cite{Kirton2019} realized in ultracold-atom cavity QED~\cite{Baumann2010,Klinder2015,Dogra2019}, circuit QED~\cite{Baksic2014} and trapped ions~\cite{SafaviNaini2018}. When two spin ensembles couple to a cavity mode with opposite signs one obtains the unconventional Dicke model (UDM)~\cite{Mivehvar2024}, with ferromagnetic (xFo) and ferrimagnetic (xFi) SR phases whose competition supports nonstationary dynamics.

Quantum synchronization describes the coordinated dynamics of quantum oscillators subject to fluctuations, dissipation and drives~\cite{Li2025,Vaidya2024,Parra2020,Xu2014,Sonar2018,Murtadho2023}. It can appear as a nonstationary limit cycle with a time-periodic density matrix~\cite{Tindall2020}, as when one-time averages oscillate in the mean-field dynamics of open Dicke models~\cite{Keeling2010,Bhaseen2012}, or within a unique, time-independent stationary state of mutually synchronized subsystems~\cite{Roulet2018}. The system studied here realizes the latter: no one-time average oscillates and the synchronized oscillation appears only in two-time correlations and in the Liouvillian spectrum. Synchronization has almost always been diagnosed through symmetry-broken observables---a finite coherent amplitude~\cite{Walter2014,Lorch2016}, quantum correlations~\cite{Roulet2018,Mari2013,Ameri2015,Bastidas2010} or spectral coherence~\cite{Bastidas2015}---and observed with trapped ions~\cite{Li2025}, spin-1 atoms~\cite{Laskar2020}, nuclear spins~\cite{Krithika2022} and circuit QED~\cite{Nigg2018}. When a discrete symmetry of the Liouvillian survives the transition, one-point order parameters vanish identically and the characterization of the synchronized phase is largely open.

Phase-sensitive control of the cavity is known to reshape Dicke superradiance: squeezed reservoirs modify collective emission and SR thresholds~\cite{Hassan1982,Agarwal1990}, and intracavity parametric amplification induces a symmetry-breaking SR transition in the single-ensemble Dicke model~\cite{Zhu2020}. With two ensembles, the parametric phase does more than shift a static threshold: it removes \emph{both} static instabilities over a finite window, leaving an oscillatory instability that synchronizes the ensembles. We show that (i) the coupling phase becomes a physical control parameter $\psi$; (ii) for a parametric gain within a band fixed by the cavity alone [Eq.~\eqref{eq:range}], a $\psi$ window opens where the only possible instability of the normal state is a Hopf bifurcation, for arbitrary spin parameters; (iii) above the Hopf threshold the ensembles phase-lock inside this window; and (iv) in the quantum steady state the oscillation is carried by a slow, parity-odd Liouvillian mode whose quality factor grows without bound with ensemble size above threshold and saturates below it. Consequently, synchronization emerges with  increasing system size, although every order parameter vanishes.

\textit{Model.}---We consider the UDM with a phase-rotated coupling quadrature and a degenerate parametric amplifier (DPA) acting on the cavity,
\begin{equation}
\hat H=\omega_c\hat a^\dagger\hat a+\omega_a(\hat S_{1z}+\hat S_{2z})+\lambda\hat X_\varphi\Delta\hat S_x+\hat H_p,
\label{eq:H}
\end{equation}
where $\hat a$ annihilates a photon of the cavity mode, $\hat S_\ell$ ($\ell=1,2$) are collective spins of size $N_\ell/2$, $\hat X_\varphi=\hat a^\dagger e^{-i\varphi}+\hat a e^{i\varphi}$ is the cavity quadrature selected by the coupling phase $\varphi$, $\Delta\hat S_x=\hat S_{1x}-\hat S_{2x}$, and $\lambda$ is the single-spin light--matter coupling, of opposite sign for the two ensembles. Here, $\omega_c$ is the cavity detuning from $\omega_p/2$ and $\omega_a$ the effective spin splitting, both defined in the frame rotating at half the frequency $\omega_p$ of the parametric pump, in which $\hat H_p$ is time independent. In a transversely pumped realization the coupling is stationary in the frame of the transverse pump laser, $\omega_L$, so the parametric pump must be phase locked at $\omega_p=2\omega_L$. The DPA term reads
\begin{equation}
\hat H_p=-\frac{i\kappa\mu}{2}\left(e^{-i\theta}\hat a^{\dagger2}-e^{i\theta}\hat a^{2}\right),
\label{eq:Hp}
\end{equation}
with parametric gain $\kappa\mu$, pump phase $\theta$ and cavity loss $2\kappa\mathcal D[\hat a]$, $\kappa$ being the field-amplitude decay rate; we parametrize the gain as $\mu=\tfrac12\sinh 2r$. Each ensemble undergoes collective decay $\Gamma_\ell\mathcal D[\hat S^-_\ell]$ and dephasing $\gamma_\phi\mathcal D[\hat S_{\ell z}]$, which preserve the Dicke subspace and damp $\hat S^-_\ell$ at the transverse rate $\gamma_\perp=\Gamma_\ell N_\ell/2+\gamma_\phi/2$. We take $\Gamma_\ell=(2\gamma_\perp-\gamma_\phi)/N_\ell$, so that both ensembles share the same $\gamma_\perp$~\cite{SM}.

In the rotated frame $\beta=\avg{\hat a}e^{i\varphi}$, the mean-field cavity equation reads
\begin{equation}
\dot\beta=-(\kappa+i\omega_c)\beta-\kappa\mu\,e^{-i\psi}\beta^*-i\lambda\Delta S_x,
\label{eq:beta}
\end{equation}
with $\psi=\theta-2\varphi$. The phase-sensitive term arises from $\hat H_p$; the dissipator contributes only $-\kappa\beta$, even for a squeezed-vacuum input (End Matter, Appendix~A). Without the DPA, $\varphi$ is a gauge redundancy; $\hat H_p$ breaks the U(1) covariance and makes $\psi$ physical. We set $\omega_c=\omega_a=\kappa=1$.
\begin{figure}[t]
\centering
\includegraphics[width=\columnwidth]{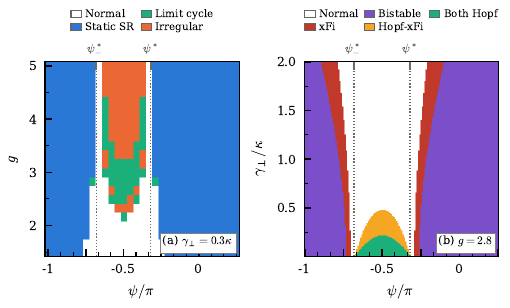}
\caption{Phase diagram ($N_2/N_1=0.4$, $\mu=1.188$). (a) $(\psi/\pi,g)$ plane at $\gamma_\perp=0.3\kappa$ from integration of the mean-field equations with collective spin decay and long-time classification: Normal; static SR; Limit cycle (regular oscillation born of the Hopf bifurcation); Irregular (quasiperiodic or chaotic). (b) $(\psi/\pi,\gamma_\perp)$ plane at $g=2.8$ from the linear analysis: Normal; static xFi; Bistable (xFi and xFo); Hopf-xFi; Both Hopf. Dotted lines: window edges $\psi^\ast_-=-\pi-\psi^\ast=-0.681\pi$ and $\psi^\ast=-0.319\pi$ (top labels).}
\label{fig:phase}
\end{figure}

\textit{$\mathbb{Z}_2$ theorem.}---Under the parity $\hat\Pi=\exp[i\pi(\hat a^\dagger\hat a+\hat S_{1z}+\hat S_{2z}+(N_1+N_2)/2)]$, $\hat H$ (including $\hat H_p$) is even and all jump operators are odd or even, so the Liouvillian commutes with $\mathcal P[\rho]=\hat\Pi\rho\hat\Pi^\dagger$~\cite{SM}. With a unique steady state (single zero eigenvalue, verified numerically) one has $\hat\Pi\rho_\infty\hat\Pi^\dagger=\rho_\infty$, hence
\begin{equation}
\avg{\hat O}_\infty=0\quad\text{for any odd-parity operator }\hat O.
\end{equation}
In particular, $\avg{\hat a}_\infty=\avg{\hat S_{\ell x}}_\infty=\avg{\hat S_{\ell y}}_\infty=0$. Products of two odd operators, such as $\avg{\hat S_{1x}\hat S_{2x}}$ or the two-time correlator $\avg{\hat S_{1x}(t+\tau)\hat S_{2x}(t)}$, are parity even and generically nonzero. Parity superselection makes $\rho_{12}=\mathrm{Tr}_{\rm cav}\rho_\infty$ block diagonal but does not imply a positive partial transpose. Numerically, the logarithmic negativity~\cite{Vidal2002,Plenio2005} vanishes for all computed states, $E_N(\rho_{12})<10^{-9}$. Since PPT states may be bound entangled~\cite{Horodecki1998}, we make no separability claim; none of the witnesses below relies on entanglement.

\textit{Instability structure.}---Linearizing about the normal state $\ket{\downarrow\downarrow}\otimes\ket{0}$ and Laplace transforming gives the characteristic equation~\cite{SM}
\begin{equation}
\big[(s+\kappa)^2+\Omega_c^2\big]\big[(s+\gamma_\perp)^2+\omega_a^2\big]=F(\psi)\,\lambda^2\omega_aN_{\rm eff},
\label{eq:char}
\end{equation}
where $\Omega_c^2=\omega_c^2-\kappa^2\mu^2$, $F(\psi)=\omega_c+\kappa\mu\sin\psi$ and $N_{\rm eff}=N_1+N_2$ for the xFi branch. The xFo branch, $N_{\rm eff}=N_1-N_2$, follows from linearizing instead about the mean-field fixed point $\ket{\downarrow\uparrow}$ (End Matter, Appendix~A). Here $s$ is the Laplace variable: perturbations evolve as $e^{st}$, and stability is lost when a root crosses the imaginary axis. A real root crossing at $s=0$ signals a static instability towards a new stationary (SR) state; a complex pair crossing at $s=\pm i\Omega$ signals a Hopf instability towards a limit cycle of frequency $\Omega$. Setting $s=0$ gives the static threshold
\begin{equation}
\lambda_{c,{\rm st}}^2=\frac{(\gamma_\perp^2+\omega_a^2)(\kappa^2+\Omega_c^2)}{\omega_aN_{\rm eff}F(\psi)},
\label{eq:st}
\end{equation}
and setting $s=i\Omega$ gives the $\psi$-independent Hopf frequency and the Hopf threshold,
\begin{align}
\Omega^2&=\frac{\gamma_\perp(\kappa^2+\Omega_c^2)+\kappa(\gamma_\perp^2+\omega_a^2)}{\kappa+\gamma_\perp},\label{eq:Om}\\
\lambda_{c,{\rm Hopf}}^2&=\frac{A_cA_d-4\kappa\gamma_\perp\Omega^2}{\omega_aN_{\rm eff}F(\psi)},\label{eq:hopf}
\end{align}
with $A_c=\kappa^2-\Omega^2+\Omega_c^2$ and $A_d=\gamma_\perp^2-\Omega^2+\omega_a^2$.

\begin{figure*}[t]
\centering
\includegraphics[width=\textwidth]{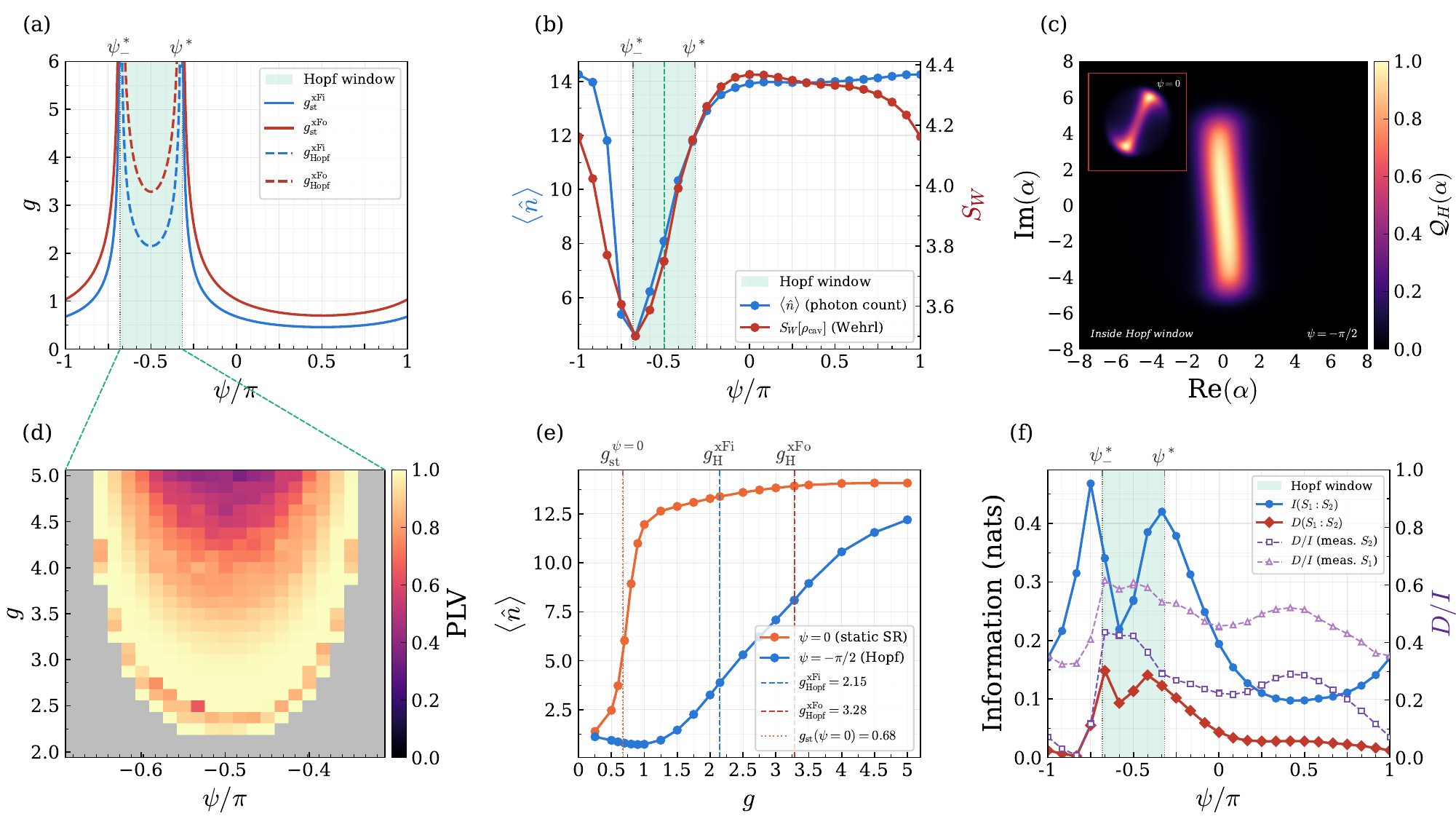}
\caption{Thresholds, synchronization map and steady-state observables ($N_2/N_1=0.4$, $\gamma_\perp=0.3\kappa$, $\mu=1.188$; $N_1=10$, $N_2=4$, $N_{\rm cav}=30$). (a) Analytical static (solid) and Hopf (dashed) thresholds; the static branches diverge at the window edges $\psi^\ast=-0.319\pi$ and $\psi^\ast_-=-\pi-\psi^\ast=-0.681\pi$ (green band, dotted lines). (b) $\avg{\hat n}(\psi)$ and Wehrl entropy $S_W[\rho_{\rm cav}](\psi)$ at $g=3.28$. (c) Husimi $\mathcal Q_H(\alpha)$ at $\psi=-\pi/2$ (inset: $\psi=0$). (d) Mean-field phase-locking value ${\rm PLV}=|\avg{e^{i(\phi_1-\phi_2)}}|$ inside the window (gray: no oscillation). (e) $\avg{\hat n}(g)$ at $\psi=-\pi/2$ and $\psi=0$; dashed lines mark $g^{\rm xFi}_{\rm Hopf}=2.15$ and $g^{\rm xFo}_{\rm Hopf}=3.28$.
At $\psi=0$, $\avg{\hat n}$ rises steeply around the mean-field static threshold $g_{\rm st}^{\psi=0}\simeq0.68$. (f) Mutual information $I$ and quantum discord $D$ (measurement on $S_2$) between the ensembles versus $\psi$ at $g=3.28$; right axis: quantum fraction $D/I$ for measurement on $S_2$ and on $S_1$.}
\label{fig:thresholds}
\end{figure*}

Both thresholds share the factor $F(\psi)^{-1}$, so their ratio is $\psi$ independent and they never cross. The window arises from a sign mismatch. Provided the cavity mode is stable ($\kappa^2+\Omega_c^2>0$), the static numerator is positive. The Hopf condition $\kappa A_d+\gamma_\perp A_c=0$ implies $A_cA_d=-\gamma_\perp A_c^2/\kappa\le0$, so the Hopf numerator is negative for $\gamma_\perp>0$. Where $F(\psi)<0$, Eq.~\eqref{eq:st} has no positive solution while Eq.~\eqref{eq:hopf} stays finite; where $F>0$ the roles reverse. The window edges are the zeros of $F$, $\psi^*=-\arcsin[\omega_c/(\kappa\mu)]$ and $\psi^*_-\equiv-\pi-\psi^*$, where the static branches diverge [Fig.~\ref{fig:thresholds}(a)]. As $\gamma_\perp\to0$ the Hopf numerator vanishes linearly, so $\lambda_{c,{\rm Hopf}}\propto\sqrt{\gamma_\perp}\to0$ and $\Omega\to\omega_a$: without spin damping, the normal state inside the window is destabilized by arbitrarily weak coupling.

\textit{Range of validity and generality.}---The window requires $\min_\psi F=\omega_c-\kappa\mu<0$, and the cavity mode of Eq.~\eqref{eq:beta} must itself be stable, $\kappa^2+\Omega_c^2>0$. Hence
\begin{equation}
\omega_c<\kappa\mu<\sqrt{\omega_c^2+\kappa^2},
\label{eq:range}
\end{equation}
which for $\omega_c=\kappa$ corresponds to $0.722<r<0.881$. At the lower edge, $\kappa\mu=\omega_c$, the window shrinks to the single point $\psi=-\pi/2$. Both bounds depend only on $\omega_c/\kappa$ and the window edges only on $\omega_c/(\kappa\mu)$, so the range can be placed wherever a given platform operates. Inside the window, the normal state can lose stability only through a Hopf bifurcation for arbitrary couplings $\lambda_\ell$ (of either sign), spin frequencies $\omega_\ell$, damping rates $\gamma_{\perp,\ell}$ and ensemble sizes $N_\ell$ (End Matter, Appendix~A).

We use $g\equiv\sqrt{N_1}\lambda/\kappa$, $\mu=1.188$ ($r=0.8$), $\gamma_\perp=0.3\kappa$ and $N_2/N_1=0.4$ ($N_1=10$, $N_2=4$). The window is then $\psi/\pi\in[-0.681,-0.319]$, with $g^{\rm xFi}_{\rm Hopf}=2.15$, $g^{\rm xFo}_{\rm Hopf}=3.28$ at $\psi=-\pi/2$ and $\Omega/2\pi=0.157\kappa$. Figure~\ref{fig:phase}(a) shows the phase structure in the $(\psi,g)$ plane from direct integration of the mean-field equations, with the collective spin decay of the model. At each $\psi$ the lowest analytic threshold coincides with the numerical normal-phase boundary. Inside the window the limit cycle is regular from $g^{\rm xFi}_{\rm Hopf}$ up to $g\approx g^{\rm xFo}_{\rm Hopf}$; above, the oscillation becomes irregular. Figure~\ref{fig:phase}(b), from the linear analysis at $g=2.8$, shows that the Hopf-only region persists up to $\gamma_\perp\approx0.48\kappa$.

\textit{Mean-field synchronization.}---We define the oscillation phases $\phi_1=\arg(S_{1x}+iS_{1y})$ and $\phi_2=\arg(-S_{2x}-iS_{2y})$ (ensemble~2 couples with the opposite sign) and quantify locking by the phase-locking value ${\rm PLV}=|\avg{e^{i(\phi_1-\phi_2)}}_t|$, which equals one for any constant phase difference. In the regular limit-cycle region the two ensembles oscillate at the same frequency, $f_1/f_2=1$, and are strictly phase locked, ${\rm PLV}\simeq1$, with a constant offset $\phi_1-\phi_2\approx(0.33$--$0.44)\pi$ that grows with $g$ (SM~\S XIII; PLV in Fig.~\ref{fig:thresholds}(d)). The Kuramoto order parameter $|\tfrac12\sum_\ell e^{i\phi_\ell}|$ is therefore below one ($\approx0.77$--$0.87$) although the locking is strict. Above $g\approx g^{\rm xFo}_{\rm Hopf}$ the dominant frequencies remain equal but the locking degrades (${\rm PLV}=0.87$ at $g=3.5$, $0.44$ at $g=5$). Detuning the ensembles yields an Arnold tongue that widens with $g$ and collapses when the shared cavity is removed, confirming cavity-mediated locking~\cite{SM,Pikovsky2003}. The staggered coupling sign only fixes the locked relative phase (End Matter, Appendix~B).

\textit{Two-time signatures in the quantum steady state.}---Because odd averages vanish, any oscillation must be read from two-time correlations. We compute $G_{AB}(\tau)=\avg{\hat A(t+\tau)\hat B(t)}_\infty=\mathrm{Tr}[\hat A\,e^{\mathcal L\tau}(\hat B\rho_\infty)]$ for $\hat A,\hat B\in\{\hat S_{1x},\hat S_{2x}\}$, which evolve in the parity-odd sector of the Liouvillian $\mathcal L$, and from them the spectra $P_{AB}(\omega)=\int d\tau\,e^{-i\omega\tau}G_{AB}(\tau)$ and the cross-spectral coherence
\begin{equation}
\mathcal C(\omega)=\frac{|P_{12}(\omega)|}{\sqrt{P_1(\omega)P_2(\omega)}}\le1 .
\label{eq:coh}
\end{equation}
Harmonic inversion of $G_{AB}(\tau)$ yields the dominant oscillatory Liouvillian eigenvalue $-\Gamma_{\rm lc}+i\Omega_q$~\cite{Minganti2018,Iemini2018}; its quality factor $Q=\Omega_q/2\Gamma_{\rm lc}$ measures how long the oscillation stays phase coherent. A mean-field limit cycle corresponds to $Q\to\infty$; in a finite system quantum noise makes its phase diffuse, $\Gamma_{\rm lc}\propto1/N$, so that $Q$ grows linearly with $N$ above the Hopf threshold. Below threshold the slow mode is a damped linear excitation of the normal state, whose damping is independent of $N$ at fixed $g$. The scaling of $Q$ with $N$ is therefore the discriminant between synchronization and a common-mode response.

\begin{figure}[t]
\centering
\includegraphics[width=\columnwidth]{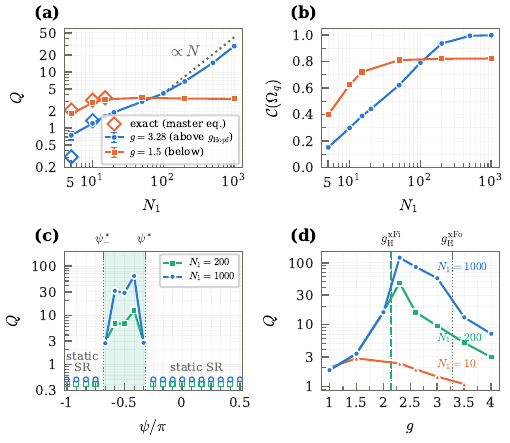}
\caption{Emergence of quantum synchronization with system size (TWA, $N_2/N_1=0.4$, $\mu=1.188$, $\gamma_\perp=0.3\kappa$; error bars: standard deviation over three noise realizations). (a) Quality factor $Q=\Omega_q/2\Gamma_{\rm lc}$ of the slow parity-odd mode at $\psi=-\pi/2$ above ($g=3.28$, circles) and below ($g=1.5$, squares) the Hopf threshold;  open diamonds: exact master-equation results ($N_1=10$ above threshold: $N_{\rm cav}=40$, and $N_1=15$ below threshold: $N_{\rm cav}=20$; lower bounds); dotted line: $Q\propto N$. (b) Cross-spectral coherence $\mathcal C(\Omega_q)$. (c) $Q$ versus $\psi$ at $g=3.28$ for $N_1=200$ and $1000$; open symbols at the bottom mark static superradiant states (no oscillatory mode); green band and dotted lines: analytical window, edges $\psi^\ast_-$ and $\psi^\ast$. (d) $Q$ versus $g$ at $\psi=-\pi/2$; dashed line: $g^{\rm xFi}_{\rm Hopf}$; dotted line: $g^{\rm xFo}_{\rm Hopf}$.}
\label{fig:emergence}
\end{figure}

\textit{Exact results.}---Full master-equation simulations at $g=3.28$ (SM~\S VI), for $N_1=10$, $N_2=4$, show that the window selects the oscillatory mode: inside it $Q=0.77$--$1.16$ and $\Omega_q=(0.84$--$1.30)\kappa$, close to $\Omega=0.99\kappa$, with $\mathcal C(\Omega_q)\approx0.4$--$0.5$. To the right $Q$ drops to $0.21$--$0.54$ and at $\psi/\pi=0.5$ no oscillatory mode remains; to the left a mode with $Q\approx1$ persists at $\psi/\pi=-0.92$ but with low coherence ($\mathcal C=0.15$), a finite-size asymmetry that disappears at large $N$ [Fig.~\ref{fig:emergence}(c)]. Above threshold these values are lower bounds ($Q=1.11\to1.37$ for $N_{\rm cav}=30\to40$). At $\psi=-\pi/2$, however, $Q$ is largest just below threshold ($Q=3.17$ at $g=1.5$, with $\mathcal C=0.89$) and decreases smoothly through $g^{\rm xFi}_{\rm Hopf}$, a signature of critical slowing down of the linear mode that at fixed small $N$ cannot be separated from synchronization.

\textit{Emergence with system size.}---To reach large $N$ we use the truncated Wigner approximation (TWA) for the cavity and the two collective spins, with stochastic terms describing cavity loss and collective decay~\cite{Polkovnikov2010,Huber2021}. The TWA reproduces the exact quality factors to within $10$--$15\%$ below threshold for $N_1=5$ and $10$, including their dependence on $g$ and $N$. At $N_1=10$ above threshold the TWA value lies within the range spanned by the unconverged exact results (SM~\S\S VI--VII). Above the Hopf threshold $Q$ grows without bound, from $1.22\pm0.02$ at $N_1=10$ to $29\pm2$ at $N_1=1000$, approaching $Q\propto N$ (local exponent $1.0$ between $N_1=500$ and $1000$), and $\mathcal C(\Omega_q)$ rises from $0.30$ to $0.999$, exceeding $0.9$ for $N_1\gtrsim200$ [Fig.~\ref{fig:emergence}(a),(b)]. Below threshold both saturate, $Q\simeq3.4$ and $\mathcal C\simeq0.82$ for $N_1\ge50$. At $N_1=1000$ an oscillatory mode exists only inside the analytical window, $\psi/\pi\in[-0.68,-0.32]$, while every point outside it is a static superradiant state [Fig.~\ref{fig:emergence}(c)]. Likewise the curves $Q(g)$ for $N_1=200$ and $1000$ coincide below $g^{\rm xFi}_{\rm Hopf}$ and separate above it [Fig.~\ref{fig:emergence}(d)]. The same scaling holds at $g=2.8$, inside the regular mean-field limit cycle ($Q=70\pm4$ at $N_1=1000$, $348$ at $N_1=5000$), and the relative phase of the two ensembles grows from $\approx0$ for $N_1\le10$ to its mean-field value, $0.373\pi$, for $N_1\ge1000$ (End Matter, Appendix~E): the slow quantum mode acquires both the coherence and the locked phase of the classical limit cycle.

\textit{Quantum signatures.}---We solve the full master equation with Hamiltonian~\eqref{eq:H} (SM~\S I) for $N_1=10$, $N_2=4$ and $N_{\rm cav}=30$~\cite{SM}. Above threshold the photon number is not converged in the cutoff ($\avg{\hat n}=6.58$, $8.09$, $9.25$ for $N_{\rm cav}=20$, $30$, $40$ at $\psi=-\pi/2$, against $15.7$ in the TWA; SM~\S VI), so exact photon numbers are lower bounds and the conclusions rest on the quality factors and the TWA scaling. The Liouvillian has a single zero eigenvalue at all reported parameters, so $\rho_\infty$ is unique. Convergence is documented in the SM; we verified $|\avg{\hat a}|<10^{-10}$, confirming the $\mathbb Z_2$ theorem to numerical precision.

\textit{Photonic witnesses.}---The photon number is strongly suppressed inside the window: at $N_1=10$, $\avg{\hat n}$ drops by 68\% and $S_W$ by 20\% (lower bounds, growing with the cutoff), both minimal at $\psi/\pi=-0.67$ [Fig.~\ref{fig:thresholds}(b)], and in the TWA at $N_1=1000$ the limit-cycle states inside the window carry one to two orders of magnitude fewer photons than the static superradiant states outside it. The Husimi function $\mathcal Q_H(\alpha)$ [Fig.~\ref{fig:thresholds}(c)] changes from two lobes (static SR, $\psi=0$) to a single elongated stripe inside the window. All three are controlled by the cavity response $F(\psi)$, which also fixes the static cavity-mediated coupling $J_{\rm eff}\propto F(\psi)$ (End Matter, Appendix~D), and persist for a single ensemble: a master-equation simulation with $N_2=0$ gives drops of 79\% and 23\% (SM Fig.~S10). They therefore certify the engineered oscillatory regime, the precondition for synchronization, and not synchronization itself. The $\psi=-\pi/2$ curve in Fig.~\ref{fig:thresholds}(e) shows changes of slope near both analytical Hopf thresholds.

\textit{Inter-ensemble correlations.}---We quantify spin--spin correlations through the mutual information $I(S_1{:}S_2)$ and its split into classical correlation $J$ and discord $D$~\cite{Ollivier2001,Henderson2001,Modi2012,MeloLuna2017}, $I=J+D$, with $D({\rm meas}\,S_\ell)$ denoting the measured subsystem. The amount of correlation does not single out the window: $I$ is largest just outside its left edge ($\psi/\pi\approx-0.75$) and at its right edge, with a minimum inside [Fig.~\ref{fig:thresholds}(f)]. What distinguishes the window is the quantum fraction $D/I$. Inside it, $D/I\simeq0.42$--$0.44$ (measuring $S_2$) and $0.59$--$0.62$ (measuring $S_1$) over most of the window, decreasing to $0.29$ ($0.54$) at its right edge, against 0.01--0.29 (0.32--0.54) in the static superradiant phases. This contrast is robust: from $N_{\rm cav}=30$ to $40$, $D/I$ changes by less than $5\%$ inside the window but decreases outside it, whereas $I$ and $D$ separately are cutoff sensitive (SM~\S X, Fig.~S12). Versus $g$ at $\psi=-\pi/2$ it decreases from $\approx0.9$ at weak coupling, where the correlations are small but mostly quantum, to $\approx0.42$ between the two Hopf thresholds ($2.5\lesssim g\lesssim3.5$) and keeps decreasing at stronger coupling, whereas at $\psi=0$ it drops to $\approx0.27$ as soon as the static threshold is crossed (SM \S\S XI--XII). The asymmetry $\Delta D=D({\rm meas}\,S_1)-D({\rm meas}\,S_2)\ge0$ (within the optimizer noise) reflects $N_1>N_2$. Since $I$ and $D$ peak below threshold ($g\approx1.5$), these one-time correlations characterize $\rho_\infty$, and the window as the region where its quantum share is largest, but do not certify synchronization; that role belongs to the finite-size scaling of Fig.~\ref{fig:emergence}.

\textit{Witness hierarchy.}---Each level below is parity even and measurable independently. The levels separate what is engineered by the cavity response from what requires two ensembles.
\begin{table}[h]
\centering
\small\setlength{\tabcolsep}{3pt}\renewcommand{\arraystretch}{1.1}
\begin{tabular}{@{}p{0.19\columnwidth}p{0.45\columnwidth}p{0.30\columnwidth}@{}}
\hline\hline
Level & Observable & Certifies\\
\hline
Photonic & $\avg{\hat n}(\psi)$, $S_W(\psi)$,~Husimi~lobes & oscillatory regime (one ensemble suffices)\\
Symmetry & $\avg{\hat a}_\infty=0$ (theorem) & $\mathbb Z_2$ protection\\
Correlations & $I(S_1{:}S_2)$, $D/I$ (largest in the window), $\Delta D\ge0$ & correlation structure of $\rho_\infty$\\
Two-time & slow odd mode $-\Gamma_{\rm lc}\pm i\Omega_q$: $Q\propto N$, $\mathcal C\to1$ above threshold; saturation below & synchronization (emergent with $N$)\\
Mean field & ${\rm PLV}=1$, $f_1=f_2$ & locking ($N\to\infty$)\\
\hline\hline
\end{tabular}
\end{table}

\textit{Experimental realization.}---The two ingredients are collective coupling of two spin ensembles to a bosonic mode with a tunable sign, and intracavity degenerate parametric amplification of that mode. In cavity QED with two condensates~\cite{Baumann2010,Mivehvar2024,Muniz2020}, pump amplitudes set $\lambda_1=-\lambda_2$ and a $\chi^{(2)}$ element or modulation of the cavity at twice the transverse-pump frequency provides $\hat H_p$, phase locked to the transverse pump, with $\psi$ tuned through the relative phase of the two pumps. The operating conditions are Eq.~\eqref{eq:range}, $\gamma_\perp\lesssim0.5\kappa$ and $g\approx2$--3. For $\kappa/2\pi\sim150$--$300$~kHz~\cite{Dogra2019,Muniz2020}, $\Omega/2\pi\approx0.16\kappa$ lies well within homodyne and heterodyne bandwidths~\cite{Brennecke2008,Mottl2012}. In trapped ions, a spin-dependent force realizes the collective coupling of one ensemble to a motional mode~\cite{SafaviNaini2018} and parametric modulation of the trap provides $\hat H_p$~\cite{Burd2019,Burd2021}; the second ensemble with opposite sign would require a second species or site-resolved control of the force, not yet demonstrated. Superconducting circuits offer flux-pumped resonators with direct control of the pump phase (End Matter, Appendix~C). The photonic witnesses require photon counting versus pump phase and heterodyne tomography, the two-time coherence heterodyne detection of the cavity output or spin-correlation spectroscopy, and the correlation level collective spin measurements of both ensembles~\cite{Hosten2016}.

\textit{Summary.}---The gain band of Eq.~\eqref{eq:range} turns the parametric phase into a switch between static superradiance and a Hopf-only regime in which the two ensembles lock. What is quantum is not the limit cycle itself, which is that of mean-field theory, but the state that carries it: a unique, parity-symmetric steady state in which no one-time observable oscillates, synchronization is encoded only in two-time correlations, and its coherence time is set by quantum phase diffusion, $\Gamma_{\rm lc}\propto1/N$. The finite-size scaling of $Q$, rather than an order parameter or an entanglement measure, is therefore the signature of synchronization in such states. The approach applies to any driven-dissipative platform with a discrete Liouvillian symmetry and, as a perspective, links synchronization to boundary time crystals~\cite{Iemini2018}, here with a phase locked between two subsystems.

\begin{acknowledgments}
We thank the Norwegian Partnership Programme for Global Academic Cooperation NORPART (QTECNOS, Grant No.~2021/10436), the Department of Physics at the University of Oslo, and Universidad del Valle (Grant No.~CI~71405) for support.
\end{acknowledgments}

\section*{End Matter}
\setcounter{equation}{0}
\renewcommand{\theequation}{A\arabic{equation}}
\def\theHequation{A\arabic{equation}}

\textit{Appendix A: Generality and robustness.}---For arbitrary $\lambda_\ell$, $\omega_\ell$, $\gamma_{\perp,\ell}$ and $N_\ell$, linearization about $\ket{\downarrow\downarrow}$ gives
\begin{equation}
K(s)D_1(s)D_2(s)=F(\psi)\big[\omega_1\lambda_1^2N_1D_2(s)+\omega_2\lambda_2^2N_2D_1(s)\big],
\label{eq:A1}
\end{equation}
with $K(s)=(s+\kappa)^2+\Omega_c^2$ and $D_\ell(s)=(s+\gamma_{\perp,\ell})^2+\omega_\ell^2$. For $\omega_1=\omega_2\equiv\omega_a$, $\gamma_{\perp,1}=\gamma_{\perp,2}$ and $\lambda_1^2=\lambda_2^2\equiv\lambda^2$, Eq.~\eqref{eq:A1} equals $D(s)$ times Eq.~\eqref{eq:char} with $N_{\rm eff}=N_1+N_2$. For real $s\ge0$, $K(s)\ge\kappa^2+\Omega_c^2>0$ and $D_\ell(s)>0$, so for $F(\psi)<0$ Eq.~\eqref{eq:A1} has no real root with $s\ge0$: inside the window the normal state can lose stability only through a Hopf bifurcation, whatever the values of these parameters. Additional individual or collective dephasing and inhomogeneous damping only renormalize $\gamma_{\perp,\ell}$, and unequal frequencies enter through $\omega_\ell$; thermal photons do not enter Eq.~\eqref{eq:beta}.  Likewise, a squeezed-vacuum input, $2\kappa\mathcal D[\hat L]$ with $\hat L=\cosh\zeta\,\hat a+e^{i\vartheta}\sinh\zeta\,\hat a^\dagger$, changes only the noise: for any $\zeta$ and $\vartheta$ its contributions in $\beta^*$ cancel and Eq.~\eqref{eq:beta} is unchanged (SM~\S II).

The thresholds depend on $\lambda$ and $N_\ell$ only through $g$ and $N_2/N_1$ (at fixed $\gamma_\perp$). The xFo branch, $N_{\rm eff}=N_1-N_2$, follows from linearizing about $\ket{\downarrow\uparrow}$, a mean-field fixed point with ensemble~2 inverted, whose response to the cavity changes sign. Under collective decay this configuration is itself unstable, so $g^{\rm xFo}$ is a threshold of the mean-field dynamics, not an instability of the normal state. Inside the window the higher, diverging static branches are linear thresholds of a state that is no longer the attractor, so the internal phase boundaries are set by nonlinear dynamics. At $\mu=1.188$ the window exists for gain variations of $-16\%$ to $+19\%$; at $g\approx3$ the synchronized region spans pump-phase deviations of about $\pm0.1\pi$ around $\psi=-\pi/2$ and persists up to $\gamma_\perp\approx0.5\kappa$ (SM~\S XIV). Full master-equation runs at $\psi=-\pi/2$, $g=3.28$ with thermal photons ($n_{\rm th}=0.1$) or collective dephasing ($\gamma_\phi=0.1\kappa$) give $Q=0.99$ and $1.13$, against $1.11$ without them (SM Table~S2).

\textit{Appendix B: One ensemble, coupling signs.}---For $N_2=0$, Eq.~\eqref{eq:A1} reduces to $K(s)D_1(s)=F(\psi)\omega_1\lambda_1^2N_1$: the window and the Hopf-only instability persist, and the photonic witnesses, controlled by $F(\psi)$, certify the oscillatory regime. Synchronization is undefined for one ensemble. The inter-ensemble quantities---$I(S_1{:}S_2)$, $J$, $D$, $\Delta D$ and $\mathcal C(\omega)$---require two ensembles. The rotation $e^{i\pi\hat S_{2z}}$ maps $\lambda_2\to-\lambda_2$ and $\hat S_2^-\to-\hat S_2^-$ and leaves the master equation invariant, so for $\lambda_2=+\lambda_1$ all witnesses coincide with those of the UDM and only the sign of $\avg{\hat S_1^+\hat S_2^-}$ changes. For $|\lambda_2|\neq|\lambda_1|$ the window and instability type are unchanged, since Eq.~\eqref{eq:A1} depends on the couplings only through $\lambda_\ell^2$.

\textit{Appendix C: Platforms and observables.}---Collective couplings of two ensembles to a bosonic mode with tunable sign, plus degenerate parametric amplification of that mode, are available in cavity QED with quantum gases~\cite{Zhu2020,Baumann2010} and in superconducting circuits with flux-pumped resonators. In trapped ions, spin-dependent forces and parametric trap modulation are demonstrated~\cite{SafaviNaini2018,Burd2019,Burd2021}, while a second, oppositely coupled ensemble remains to be implemented.

\textit{Appendix D: Effective coupling.}---The static cavity response ($s=0$) of Eq.~\eqref{eq:beta} yields the cavity-mediated spin--spin coupling~\cite{SM}
\begin{equation}
J_{\rm eff}(\psi)=\lambda^2\frac{F(\psi)}{\kappa^2+\Omega_c^2},
\label{eq:A2}
\end{equation}
which vanishes at the window edges and is negative throughout the window. It is the exact static coupling of the mean-field equations, not a dynamical effective Hamiltonian, since $\kappa$ is not large compared with $\lambda\sqrt{N_\ell}$ and $\omega_a$; it identifies the common factor $F(\psi)$ behind the thresholds and the photonic witnesses.

\textit{Appendix E: Robustness of the emergence.}---At $g=2.8$ the TWA quality factor grows from $1.66$ at $N_1=10$ to the values quoted in the main text. The relative phase of the two ensembles is read from the cross spectrum at $\Omega_q$. For $N_1\le10$ exact and TWA values agree, and for $N_1\ge1000$ the phase coincides with the same quantity evaluated on the mean-field limit cycle (SM~\S XV).

\clearpage
\onecolumngrid

\begin{center}
{\large\bfseries Supplemental Material for\\ ``Emergent Quantum Synchronization Without Symmetry Breaking''}
\end{center}

\providecommand{\ket}[1]{|#1\rangle}
\providecommand{\avg}[1]{\langle #1\rangle}
\providecommand{\figfile}[2]{%
  \IfFileExists{#1}{\includegraphics[width=#2]{#1}}%
  {\fbox{\parbox[c][3.2cm][c]{0.92\dimexpr#2\relax}{\centering\small\texttt{\detokenize{#1}}\\(insert figure)}}}}

\setcounter{section}{0}
\setcounter{figure}{0}
\setcounter{table}{0}
\setcounter{equation}{0}
\setcounter{secnumdepth}{3}
\renewcommand{\thesection}{\Roman{section}}
\renewcommand{\thefigure}{S\arabic{figure}}
\renewcommand{\thetable}{S\arabic{table}}
\renewcommand{\theequation}{S\arabic{equation}}
\def\theHsection{S\arabic{section}}
\def\theHfigure{S\arabic{figure}}
\def\theHtable{S\arabic{table}}
\def\theHequation{S\arabic{equation}}

\noindent This Supplemental Material provides: (\S I) the model, including the degenerate parametric amplifier;
(\S II) mean-field equations and the dissipator convention; (\S III) the linear-stability analysis and the origin of
the Hopf-only window; (\S IV) the characteristic equation for general parameters; (\S V) synchronization in the unique
symmetric steady state; (\S VI) exact two-time correlations; (\S VII) the truncated Wigner approximation and the
finite-size scaling; (\S VIII) the effective coupling $J_{\rm eff}$; (\S IX) the $\mathbb Z_2$ theorem; (\S X) numerical
methods and convergence; (\S XI) quantum fraction $D/I$ of the inter-ensemble correlations; (\S XII) cavity--spin mutual information;
(\S XIII) mean-field synchronization figures; (\S XIV) parameter ranges, coupling signs and robustness; and
(\S XV) robustness of the emergence and the locked phase difference.

\section{Model}

The unconventional Dicke Hamiltonian~\cite{Mivehvar2024} with $\lambda_1=-\lambda_2\equiv\lambda$ reads
\begin{equation}
\hat H_{\rm uD}=\omega_c\hat a^\dagger\hat a+\omega_a(\hat S_{1z}+\hat S_{2z})+\lambda(\hat a^\dagger+\hat a)(\hat S_{1x}-\hat S_{2x}).
\end{equation}
It is unitarily equivalent to the standard Dicke model via $\hat U=e^{-i\pi\hat S_{2z}}$, but $\hat{\mathbf S}_1^2$ and
$\hat{\mathbf S}_2^2$ are separately conserved. We rotate the coupling quadrature by a phase $\varphi$ and
add a degenerate parametric amplifier (DPA) acting on the cavity,
\begin{align}
\hat H&=\omega_c\hat a^\dagger\hat a+\omega_a(\hat S_{1z}+\hat S_{2z})+\lambda(\hat a^\dagger e^{-i\varphi}+\hat ae^{i\varphi})(\hat S_{1x}-\hat S_{2x})+\hat H_p,\label{eqS:H}\\
\hat H_p&=-\frac{i\kappa\mu}{2}\left(e^{-i\theta}\hat a^{\dagger2}-e^{i\theta}\hat a^{2}\right),\label{eqS:Hp}
\end{align}
with parametric gain $\kappa\mu$ and pump phase $\theta$. The full master equation is
\begin{equation}
\dot\rho=-i[\hat H,\rho]+2\kappa\mathcal D[\hat a]\rho+\sum_{\ell=1,2}\left(\Gamma_\ell\mathcal D[\hat S_\ell^-]+\gamma_\phi\mathcal D[\hat S_{\ell z}]\right)\rho,
\label{eqS:ME}
\end{equation}
with $\mathcal D[\hat L]\rho=\hat L\rho\hat L^\dagger-\tfrac12\{\hat L^\dagger\hat L,\rho\}$. Without $\hat H_p$, $\varphi$
is a gauge redundancy absorbed by $\hat a\to\hat ae^{i\varphi}$: $\hat H_p$ breaks this U(1) covariance and makes
$\psi=\theta-2\varphi$ a physical control parameter. We parametrize the gain as $\mu=\tfrac12\sinh 2r$; $r$ parametrizes the gain and is not the squeezing of
any input field.

\section{Mean-field equations and dissipator convention}

In the rotated frame $\beta=\alpha e^{i\varphi}$, $\alpha=\avg{\hat a}$, the cavity equation is (main text Eq.~(3))
\begin{equation}
\dot\beta=-(\kappa+i\omega_c)\beta-\kappa\mu\,e^{-i\psi}\beta^*-i\lambda\Delta S_x .
\label{eqS:beta}
\end{equation}
The loss $2\kappa\mathcal D[\hat a]$ gives $-\kappa\alpha$, so $\kappa$ is the
field-amplitude decay rate. The phase-sensitive term comes from $\hat H_p$: $i\avg{[\hat H_p,\hat a]}=-\kappa\mu
e^{-i\theta}\alpha^*$, i.e.\ $-\kappa\mu e^{-i\psi}\beta^*$ in the rotated frame.
We note that a squeezed reservoir does not produce this term. For a squeezed-vacuum input with jump operator
$\hat L=u\hat a+v\hat a^\dagger$, $u=\cosh\zeta$, $v=e^{i\vartheta}\sinh\zeta$ (squeezing $\zeta$ and phase $\vartheta$ of the
input), and any operator $\hat O$,
\begin{equation}
\mathrm{Tr}\big[\hat O\,\mathcal D[\hat L]\rho\big]=\tfrac12\avg{[\hat L^\dagger,\hat O]\hat L+\hat L^\dagger[\hat O,\hat L]} .
\end{equation}
For $\hat O=\hat a$, $[\hat a,\hat L]=v$ and $[\hat L^\dagger,\hat a]=-u^*$, so
\begin{equation}
\dot\alpha|_{\rm diss}=\frac{\kappa'}{2}\avg{v\hat L^\dagger-u^*\hat L}=-\frac{\kappa'}{2}(|u|^2-|v|^2)\alpha=-\frac{\kappa'}{2}\alpha
\end{equation}
for a dissipator $\kappa'\mathcal D[\hat L]$: the terms in $\alpha^*$ cancel exactly, for any $\zeta$ and $\vartheta$. A squeezed reservoir therefore
changes only the noise and leaves Eq.~\eqref{eqS:beta} unchanged; the phase-sensitive gain must be coherent.

\textit{Spin damping.} Linearized about the polarized state $\ket{\downarrow}$, collective decay damps
$\hat S^-_\ell$ at rate $\Gamma_\ell N_\ell/2$ and dephasing at rate $\gamma_\phi/2$, so the transverse rate is
$\gamma_\perp=\Gamma_\ell N_\ell/2+\gamma_\phi/2$. We choose $\Gamma_\ell=(2\gamma_\perp-\gamma_\phi)/N_\ell$, so both
ensembles share the same $\gamma_\perp$; with $\gamma_\phi=0$, $\Gamma_\ell=2\gamma_\perp/N_\ell$. The polarized
normal state is $\alpha=0$, $S^{(0)}_{\ell z}=-\sigma_\ell N_\ell/2$, with $N_{\rm eff}=\sigma_1N_1+\sigma_2N_2$.
Linearizing about $\ket{\downarrow\downarrow}$ ($\sigma_1=\sigma_2=1$) gives the xFi branch, $N_{\rm eff}=N_1+N_2$.
The xFo branch, $N_{\rm eff}=N_1-N_2$, is obtained by linearizing about $\ket{\downarrow\uparrow}$ ($\sigma_1=1$,
$\sigma_2=-1$), which is also a fixed point of the mean-field equations. With ensemble~2 inverted, the term
$-\lambda_2XS^{(0)}_{2z}$ in $\dot S_{2y}$ changes sign, so the response of ensemble~2 to the cavity field adds with the
opposite sign to that of ensemble~1; for equal transverse damping of the two ensembles the characteristic equation is
Eq.~(5) of the main text with $N_{\rm eff}=N_1-N_2$. Under collective decay the inverted ensemble is itself unstable
(its transverse components grow at rate $\gamma_\perp$), so the xFo thresholds are thresholds of the mean-field dynamics
and do not describe an instability of the normal state $\ket{\downarrow\downarrow}$.

\section{Linear-stability analysis}

Expanding $S_{\ell\alpha}=S^{(0)}_{\ell\alpha}+\delta S_{\ell\alpha}$ to first order,
\begin{align}
\delta\dot S_{\ell x}&=-\omega_a\delta S_{\ell y}-\gamma_\perp\delta S_{\ell x},\\
\delta\dot S_{\ell y}&=+\omega_a\delta S_{\ell x}-\gamma_\perp\delta S_{\ell y}+(-1)^\ell\lambda S^{(0)}_{\ell z}X_\varphi .
\end{align}
Laplace transforming and summing over the ensembles gives the collective spin susceptibility
\begin{equation}
\Delta S_x(s)=-\lambda\omega_a\chi_s(s)X_\varphi(s),\qquad \chi_s(s)=\frac{-N_{\rm eff}/2}{(s+\gamma_\perp)^2+\omega_a^2}.
\end{equation}
Laplace transforming Eq.~\eqref{eqS:beta} for $(\beta,\beta^*)$, substituting $X_\varphi=\beta+\beta^*$ and setting the
determinant to zero yields, with $\Omega_c^2=\omega_c^2-\kappa^2\mu^2$ and $F(\psi)=\omega_c+\kappa\mu\sin\psi$,
\begin{equation}
\big[(s+\kappa)^2+\Omega_c^2\big]\big[(s+\gamma_\perp)^2+\omega_a^2\big]=F(\psi)\,\lambda^2\omega_aN_{\rm eff}.
\label{eqS:char}
\end{equation}
\textit{Meaning of $s$.} A linear perturbation of the normal state evolves as $e^{st}$, with $s$ a root of
Eq.~\eqref{eqS:char}. The normal state is stable while all roots have ${\rm Re}\,s<0$. Stability is lost through a real
root at $s=0$ (static instability towards a new stationary superradiant state) or through a complex pair at
$s=\pm i\Omega$ (Hopf instability towards a limit cycle of frequency $\Omega$).

\textit{Thresholds.} With $P\equiv\kappa^2+\Omega_c^2$ and $Q\equiv\gamma_\perp^2+\omega_a^2$,
\begin{align}
\lambda^2_{c,{\rm st}}&=\frac{PQ}{\omega_aN_{\rm eff}F(\psi)},\label{eqS:st}\\
\Omega^2&=\frac{\gamma_\perp P+\kappa Q}{\kappa+\gamma_\perp},\label{eqS:Om}\\
\lambda^2_{c,{\rm Hopf}}&=\frac{A_cA_d-4\kappa\gamma_\perp\Omega^2}{\omega_aN_{\rm eff}F(\psi)},\qquad A_c=P-\Omega^2,\ A_d=Q-\Omega^2 .\label{eqS:hopf}
\end{align}
Setting $\mu=0$ and $\gamma_\perp=0$ in Eq.~\eqref{eqS:st} recovers Ref.~\cite{Mivehvar2024}.
Equation~\eqref{eqS:Om} gives $\Omega^2>0$ for all $\gamma_\perp\ge0$; at $\gamma_\perp=0$, $\Omega=\omega_a$.

\textit{Origin of the Hopf-only window: a sign mismatch.} Both thresholds share the factor $F(\psi)^{-1}$, so
their ratio is independent of $\psi$ and they never cross; the window comes from the signs of the numerators.
For a stable cavity mode, $P>0$, so the static numerator $PQ$ is positive. The Hopf condition (imaginary part of
Eq.~\eqref{eqS:char} at $s=i\Omega$) is $\kappa A_d+\gamma_\perp A_c=0$, hence $A_cA_d=-\gamma_\perp A_c^2/\kappa\le0$
and the Hopf numerator $A_cA_d-4\kappa\gamma_\perp\Omega^2$ is negative for $\gamma_\perp>0$. Therefore, where
$F(\psi)>0$ only the static threshold is positive, and where $F(\psi)<0$ only the Hopf threshold is positive. The
window is the set $F(\psi)<0$, with edges $\psi^*=-\arcsin[\omega_c/(\kappa\mu)]$ and $\psi^*_-\equiv-\pi-\psi^*$, where the
static branch diverges. As $\gamma_\perp\to0$ the Hopf numerator vanishes linearly in $\gamma_\perp$, so
$\lambda_{c,{\rm Hopf}}\propto\sqrt{\gamma_\perp}\to0$: without spin damping the normal state inside the window is
destabilized by arbitrarily weak coupling, rather than becoming stable.

\textit{Range of validity.} The window exists if and only if $\min_\psi F=\omega_c-\kappa\mu<0$, and the cavity mode
is stable if and only if $P=\kappa^2+\omega_c^2-\kappa^2\mu^2>0$. Hence
\begin{equation}
\omega_c<\kappa\mu<\sqrt{\omega_c^2+\kappa^2},\qquad\text{i.e.}\qquad \tfrac12{\rm arcsinh}\frac{2\omega_c}{\kappa}<r<\tfrac12{\rm arcsinh}\frac{2\sqrt{\omega_c^2+\kappa^2}}{\kappa},
\label{eqS:range}
\end{equation}
which for $\omega_c=\kappa$ gives $0.722<r<0.881$ ($1<\mu<\sqrt2$). The estimate $r\gtrsim\ln2$ follows from
$\sinh2r\simeq e^{2r}/2$. At $\kappa\mu=\omega_c$ the window shrinks to the point $\psi=-\pi/2$. The bounds depend
only on $\omega_c/\kappa$ and the window edges only on $\omega_c/(\kappa\mu)$. For $\mu=1.188$ the window exists
for gain variations between $-16\%$ and $+19\%$. The thresholds depend on $\lambda$ and $N_\ell$ only through
$g=\sqrt{N_1}\lambda/\kappa$ and $N_2/N_1$ at fixed $\gamma_\perp$ (Fig.~\ref{fig:S6}).

\section{Characteristic equation for general parameters}

For arbitrary couplings $\lambda_\ell$ (of either sign), spin frequencies $\omega_\ell$, transverse rates
$\gamma_{\perp,\ell}$ and sizes $N_\ell$, linearizing about $\ket{\downarrow\downarrow}$ gives
\begin{equation}
\det(s-\mathbb{M})=K(s)D_1(s)D_2(s)-F(\psi)\big[\omega_1\lambda_1^2N_1D_2(s)+\omega_2\lambda_2^2N_2D_1(s)\big],
\label{eqS:general}
\end{equation}
with $K(s)=(s+\kappa)^2+\Omega_c^2$ and $D_\ell(s)=(s+\gamma_{\perp,\ell})^2+\omega_\ell^2$;
$\mathbb{M}$ denotes the Jacobian of the linearized equations. For
$\omega_1=\omega_2\equiv\omega_a$, $\gamma_{\perp,1}=\gamma_{\perp,2}$ and $\lambda_1^2=\lambda_2^2\equiv\lambda^2$ it
reduces to $D(s)$ times Eq.~\eqref{eqS:char} with $N_{\rm eff}=N_1+N_2$.
For real $s\ge0$, $K(s)\ge P>0$ and
$D_\ell(s)>0$; therefore, whenever $F(\psi)<0$, Eq.~\eqref{eqS:general} is strictly positive and has no real root with
$s\ge0$. Inside the window the normal state can only lose stability through a Hopf bifurcation, for all values of
these parameters. Additional dephasing and inhomogeneous damping only renormalize $\gamma_{\perp,\ell}$, and thermal
photons do not enter Eq.~\eqref{eqS:beta}.

\section{Synchronization in the unique symmetric steady state}

The $\mathbb Z_2$ theorem (\S IX) guarantees that $\rho_\infty$ is unique and parity symmetric, so all one-time
averages of odd operators vanish. Synchronization must then be discussed at two levels.

\textit{Mean-field level.} In the limit $N_1,N_2\to\infty$ at fixed ratio, the mean-field equations have a
limit-cycle attractor inside the window on which the two ensembles oscillate at the same frequency, $f_1/f_2=1$, and are
strictly phase locked (phase-locking value ${\rm PLV}\simeq1$) with a constant offset (main-text Fig.~2(d) and \S XIII). For deterministic mean-field trajectories, the spectral coherence of the two
ensembles is unity by construction once they lock; it is not, by itself, evidence of quantum synchronization.

\textit{Quantum level.} In $\rho_\infty$ the oscillation survives only in parity-even two-time correlations. We
use
\begin{equation}
G_{AB}(\tau)=\avg{\hat A(t+\tau)\hat B(t)}_\infty=\mathrm{Tr}\big[\hat A\,e^{\mathcal L\tau}(\hat B\rho_\infty)\big],\qquad
P_{AB}(\omega)=\int_{-\infty}^{\infty}d\tau\,e^{-i\omega\tau}G_{AB}(\tau),
\label{eqS:corr}
\end{equation}
for $\hat A,\hat B\in\{\hat S_{1x},\hat S_{2x}\}$, with $P_\ell\equiv P_{\ell\ell}$, and the cross-spectral coherence
\begin{equation}
\mathcal C(\omega)=\frac{|P_{12}(\omega)|}{\sqrt{P_1(\omega)P_2(\omega)}}\le1 .
\label{eqS:coh}
\end{equation}
Since $\hat B\rho_\infty$ is parity odd, $G_{AB}(\tau)$ evolves entirely in the odd sector of $\mathcal L$. Its slowest
oscillatory eigenvalue $-\Gamma_{\rm lc}+i\Omega_q$ describes the quantum remnant of the limit cycle: $\Omega_q$ is its
frequency and $\Gamma_{\rm lc}^{-1}$ its coherence time~\cite{Minganti2018,Iemini2018}. The quality factor
$Q=\Omega_q/2\Gamma_{\rm lc}$ diverges in the thermodynamic limit if the mean-field limit cycle is recovered. In a finite
system the Hopf bifurcation becomes a crossover, and two ensembles driven by the same cavity can show
$\mathcal C(\omega)$ of order one even without locking (common drive). The discriminating quantities are therefore the
growth of $Q$ inside the window and above threshold, and its scaling with $N_\ell$: above threshold the phase of
the limit cycle diffuses at a rate $\Gamma_{\rm lc}\propto1/N$, so $Q\propto N$, whereas below threshold the slow mode is a
damped linear excitation whose rate does not depend on $N$ at fixed $g$.

\textit{Photonic and correlation observables.} The observables $\avg{\hat n}(\psi)$, $S_W(\psi)$, the Husimi function and
the mutual-information/discord quantities are one-time averages of parity-even operators. They are consistent with the
$\mathbb Z_2$ theorem and characterize $\rho_\infty$, but they do not by themselves certify synchronization.

\section{Exact two-time correlations}

\textit{Method.} We integrate Eq.~\eqref{eqS:ME} directly on the density matrix with an adaptive Dormand--Prince
RK5(4) scheme, applying each operator in banded form. The steps are: (i) evolve from $\ket{0}\ket{\downarrow}\ket{\downarrow}$
for $t_{\rm ss}=150\,\kappa^{-1}$, which gives $\rho_\infty$ with residual $\|\mathcal L\rho\|/\|\rho\|\lesssim10^{-3}$;
(ii) evolve $\hat S_{\ell x}\rho_\infty$ for $\ell=1,2$ over $\tau_{\max}=60\,\kappa^{-1}$ and record $G_{AB}(\tau)$;
(iii) extract the dominant oscillatory eigenvalue by matrix-pencil harmonic inversion of $G_{11}+G_{22}$;
(iv) evaluate $P_{AB}(\omega)$ by quadrature over the computed window, completed analytically with the fitted modes.
The code was validated against exact diagonalization of the superoperator for $N_1=4$, $N_2=2$, $N_{\rm cav}=5$: the
Liouvillian agrees with an independent Kronecker-product construction to $3\times10^{-16}$, no eigenvalue has positive
real part, and the fitted frequency matches an exact eigenvalue. In every run $|\avg{\hat a}_\infty|=0$ to numerical precision.

\begin{figure}[ht]
\centering
\figfile{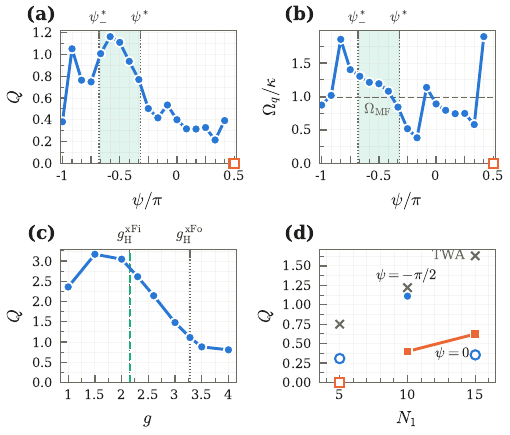}{0.62\textwidth}
\caption{Exact two-time results ($N_1=10$, $N_2=4$, $N_{\rm cav}=30$). (a) Quality factor $Q=\Omega_q/2\Gamma_{\rm lc}$ of the slow parity-odd mode and (b) its frequency $\Omega_q$ versus $\psi$ at
$g=3.28$; green band: analytical window; dashed line: mean-field $\Omega$. (c) $Q$ versus $g$ at $\psi=-\pi/2$; dashed:
$g^{\rm xFi}_{\rm Hopf}$. (d) $Q$ versus $N_1$ at $g=3.28$ ($N_2/N_1=0.4$) inside ($\psi=-\pi/2$) and outside ($\psi=0$)
the window; open squares: no oscillatory mode ($\Omega_q\simeq0$, static SR). Dotted lines mark the window edges ($\psi^\ast_-=-\pi-\psi^\ast$ and $\psi^\ast$, main text)
and the Hopf thresholds. In (d), open circles mark fits that select a mode far from the Hopf frequency
($N_1=5$ and $15$ at $\psi=-\pi/2$, cutoff limited) and crosses give the TWA values, which have no cutoff.}
\label{fig:exact}
\end{figure}

\begin{figure}[ht]
\centering
\figfile{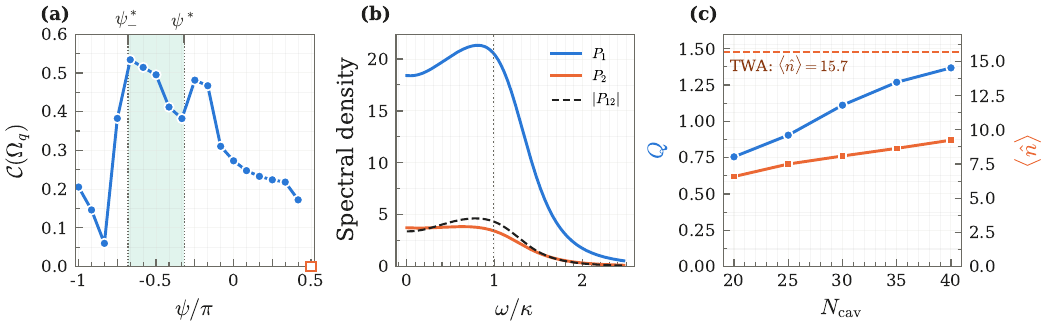}{\textwidth}
\caption{(a) Cross-spectral coherence $\mathcal C(\Omega_q)$ versus $\psi$ at $g=3.28$. (b) Spectra $P_1$, $P_2$ and
$|P_{12}|$ at $\psi=-\pi/2$; dotted line: mean-field $\Omega$. (c) Cutoff dependence at $\psi=-\pi/2$, $g=3.28$:
$Q$ (left axis) and $\avg{\hat n}$ (right axis); dashed: TWA photon number, which has no cutoff. Open square in (a):
no oscillatory mode.}
\label{fig:exactspec}
\end{figure}

\textit{Results.} Figure~\ref{fig:exact} and Tables~\ref{tab:exactpsi}--\ref{tab:exactother} report the results for
$N_1=10$, $N_2=4$, $N_{\rm cav}=30$; the cutoff dependence at $\psi=-\pi/2$ was followed up to $N_{\rm cav}=40$.
\begin{enumerate}
\item \textit{Mode selection in $\psi$.} Inside the window $Q=0.77$--$1.16$, maximal at $\psi/\pi=-0.58$, and
$\Omega_q=0.84$--$1.30\,\kappa$, close to $\Omega=0.987\kappa$. To the right of the window $Q$ drops to $0.21$--$0.54$;
at $\psi/\pi=0.417$ the fit selects a different mode ($\Omega_q=1.90\kappa$), and at $\psi/\pi=0.5$ there is no
oscillatory mode at all (static superradiance). To the left, $Q$ remains $\sim1$ at $\psi/\pi=-0.92$ but the inter-ensemble coherence is low,
$\mathcal C=0.15$, against $\mathcal C\approx0.5$ in the left half of the window. This asymmetry, absent in the linear
analysis, which depends on $\psi$ only through $\sin\psi$, disappears at large $N$ (\S VII).
\item \textit{Critical slowing down.} At $\psi=-\pi/2$, $Q$ is largest just below threshold ($Q=3.17$ at $g=1.5$,
$3.05$ at $g=2.0$) and decreases smoothly through and above it ($2.61$, $2.14$, $1.48$, $1.11$, $0.88$, $0.81$ at
$g=2.3$, $2.6$, $3.0$, $3.28$, $3.5$, $4.0$), with $\mathcal C$ decreasing from $0.92$ to $0.28$. At $N_1=10$ the
threshold leaves no abrupt signature in $Q$. The
large coherence below threshold reflects the common response of both ensembles to the cavity mode.
\item \textit{Size dependence.} Below threshold, $Q$ grows from $2.11$ at $N_1=5$ to $3.17$ at $N_1=10$ ($g=1.5$).
Above threshold the exact calculation is limited by the cavity cutoff, and increasingly so as $N_1$ grows, because
the photon number scales with $N_1$. At $N_1=15$ ($N_2=6$, dimension 3360), $g=3.28$, $\psi=-\pi/2$, the exact steady
state has $\avg{\hat n}=10.2$ against $22.3$ in the TWA, i.e.\ the photon distribution is truncated by
$N_{\rm cav}=30$, and the fit selects a mode at $\Omega_q=0.53\kappa$, far from the Hopf frequency ($Q=0.36$; TWA:
$Q=1.63$ at $\Omega_q=1.24\kappa$). The same happens at $N_1=5$ ($\Omega_q=0.41\kappa$). At $\psi=0$ the exact
$N_1=15$ state ($\avg{\hat n}=13.7$, $Q=0.62$) is likewise cutoff dominated: in the TWA it is a static superradiant
state with $\avg{\hat n}\approx5\times10^2$. Exact sizes beyond $N_1=10$ therefore do not add reliable information above
threshold at the cutoffs we can afford; the large-$N$ behavior is obtained with the TWA (\S VII). Below threshold
($g=1.5$) the photon number is small but its distribution has a long tail from the parametric drive, so even there the
cutoff matters at $N_1=15$: $\avg{\hat n}=1.40$, $1.54$ and $1.64$ for $N_{\rm cav}=12$, $15$ and $20$, converging to
the TWA value $1.65$, while $Q=2.74$, $3.15$ and $3.50$ is not yet converged; at $N_{\rm cav}=20$ it differs from the
TWA value, $3.29\pm0.04$, by $6\%$, comparable to the TWA--exact agreement at $N_1=5$ and $10$. The $N_{\rm cav}=20$ value is shown in main-text Fig.~3(a) as a lower bound.
\item \textit{Robustness.} At $\psi=-\pi/2$, $g=3.28$, thermal photons ($n_{\rm th}=0.1$) and collective dephasing
($\gamma_\phi=0.1\kappa$) leave the mode essentially unchanged ($Q=0.99$ and $1.13$, against $1.11$).
\item \textit{Cavity cutoff.} Above threshold neither the photon number nor $Q$ is converged at the largest cutoff
we can afford [Fig.~\ref{fig:exactspec}(c)]: for $N_{\rm cav}=20$, $25$, $30$, $35$, $40$ we find
$\avg{\hat n}=6.58$, $7.50$, $8.09$, $8.65$, $9.25$ (TWA, without cutoff: $15.7$) and $Q=0.76$, $0.90$, $1.11$, $1.27$,
$1.37$, with decreasing increments, while $\Omega_q$ and $\mathcal C(\Omega_q)\approx0.5$ change little. The exact
values above threshold are therefore lower bounds on $\avg{\hat n}$ and $Q$; the conclusions rely on the TWA
scaling, which has no cutoff.
\end{enumerate}

\begin{figure}[ht]
\centering
\figfile{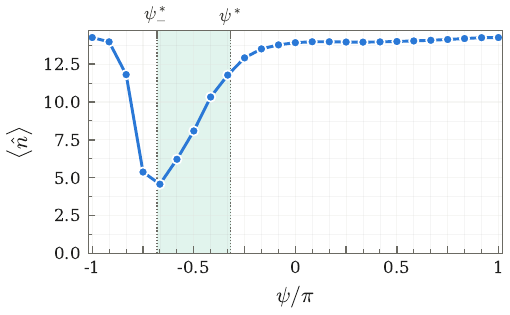}{0.5\textwidth}
\caption{Exact steady-state photon number versus $\psi$ at $g=3.28$ ($N_1=10$, $N_2=4$, $N_{\rm cav}=30$); green band:
analytical window; dotted lines: window edges $\psi^\ast_-$ and $\psi^\ast$; $\psi=\pi$ is the same point as $\psi=-\pi$. Above threshold these photon numbers are not converged in the cutoff
and are lower bounds [Fig.~\ref{fig:exactspec}(c)]; the depletion inside the window is the robust feature.}
\label{fig:nphoton}
\end{figure}

\begin{table}[ht]
\centering
\small
\begin{tabular}{rrrrrr}
\hline\hline
$\psi/\pi$ & $\avg{\hat n}$ & $\Gamma_{\rm lc}/\kappa$ & $\Omega_q/\kappa$ & $Q$ & $\mathcal C(\Omega_q)$\\
\hline
$-1.000$ & 14.25 & 1.141 & 0.872 & 0.38 & 0.21\\
$-0.917$ & 13.96 & 0.483 & 1.017 & 1.05 & 0.15\\
$-0.833$ & 11.82 & 1.214 & 1.855 & 0.76 & 0.06\\
$-0.750$ & 5.38 & 0.936 & 1.402 & 0.75 & 0.38\\
$-0.667$ & 4.58 & 0.646 & 1.303 & 1.01 & 0.53\\
$-0.583$ & 6.21 & 0.521 & 1.212 & 1.16 & 0.51\\
$-0.500$ & 8.09 & 0.536 & 1.191 & 1.11 & 0.50\\
$-0.417$ & 10.33 & 0.575 & 1.079 & 0.94 & 0.41\\
$-0.333$ & 11.77 & 0.546 & 0.841 & 0.77 & 0.38\\
$-0.250$ & 12.91 & 0.515 & 0.518 & 0.50 & 0.48\\
$-0.167$ & 13.51 & 0.458 & 0.383 & 0.42 & 0.47\\
$-0.083$ & 13.77 & 1.060 & 1.136 & 0.54 & 0.31\\
$0$ & 13.92 & 1.109 & 0.889 & 0.40 & 0.27\\
$+0.083$ & 13.98 & 1.253 & 0.793 & 0.32 & 0.25\\
$+0.167$ & 13.97 & 1.178 & 0.743 & 0.32 & 0.23\\
$+0.250$ & 13.96 & 1.136 & 0.747 & 0.33 & 0.22\\
$+0.333$ & 13.95 & 1.352 & 0.581 & 0.21 & 0.22\\
$+0.417$ & 13.97 & 2.412 & 1.896 & 0.39 & 0.17\\
$+0.500$ & 13.99 & -- & -- & -- & --\\
\hline\hline
\end{tabular}
\caption{Exact results versus $\psi$ ($g=3.28$, $N_1=10$, $N_2=4$, $N_{\rm cav}=30$). Rows $-0.667$ to $-0.333$ lie inside
the window $[-0.681,-0.319]$. At $\psi/\pi=-0.833$ and $+0.417$ the fit selects a higher-frequency mode; at $\psi/\pi=+0.5$ there is no
oscillatory mode (static SR).}
\label{tab:exactpsi}
\end{table}

\begin{table}[ht]
\centering
\small
\begin{tabular}{lrrrrr}
\hline\hline
Run ($\psi=-\pi/2$) & $\avg{\hat n}$ & $\Gamma_{\rm lc}/\kappa$ & $\Omega_q/\kappa$ & $Q$ & $\mathcal C(\Omega_q)$\\
\hline
$g=1.0$ & 0.735 & 0.214 & 1.011 & 2.36 & 0.92\\
$g=1.5$ & 1.459 & 0.165 & 1.044 & 3.17 & 0.89\\
$g=2.0$ & 3.252 & 0.179 & 1.090 & 3.05 & 0.78\\
$g=2.3$ & 4.507 & 0.214 & 1.119 & 2.61 & 0.71\\
$g=2.6$ & 5.679 & 0.266 & 1.140 & 2.14 & 0.65\\
$g=3.0$ & 7.080 & 0.390 & 1.156 & 1.48 & 0.57\\
$g=3.28$ & 8.09 & 0.536 & 1.191 & 1.11 & 0.50\\
$g=3.5$ & 8.945 & 0.703 & 1.237 & 0.88 & 0.42\\
$g=4.0$ & 10.56 & 0.912 & 1.473 & 0.81 & 0.28\\
$N_1=5$, $N_2=2$, $g=1.5$ & 1.241 & 0.260 & 1.099 & 2.11 & 0.68\\
$N_1=5$, $N_2=2$, $g=3.28$ & 5.93 & 0.673 & 0.415 & 0.31 & 0.15\\
$N_{\rm cav}=20$ & 6.58 & 0.749 & 1.131 & 0.76 & 0.39\\
$N_{\rm cav}=25$ & 7.50 & 0.636 & 1.150 & 0.90 & 0.45\\
$N_{\rm cav}=35$ & 8.65 & 0.474 & 1.202 & 1.27 & 0.50\\
$N_{\rm cav}=40$ & 9.25 & 0.448 & 1.226 & 1.37 & 0.48\\
$n_{\rm th}=0.1$ & 7.94 & 0.609 & 1.201 & 0.99 & 0.48\\
$\gamma_\phi=0.1\kappa$ & 7.75 & 0.579 & 1.304 & 1.13 & 0.32\\
\hline\hline
\end{tabular}
\caption{Exact results at $\psi=-\pi/2$: coupling scan, ensemble size, cavity cutoff and additional noise. Unless
stated, $N_1=10$, $N_2=4$, $N_{\rm cav}=30$, $g=3.28$. The $n_{\rm th}$ run has a steady-state residual of
$1.4\times10^{-2}$.}
\label{tab:exactother}
\end{table}

\section{Truncated Wigner approximation and finite-size scaling}

\textit{Stochastic equations.} We sample the Wigner function of the cavity and of the two collective spins and evolve
each sample with the Stratonovich equations~\cite{Polkovnikov2010,Huber2021}
\begin{align}
d\alpha&=\big[-(\kappa+i\omega_c)\alpha-\kappa\mu e^{-i\psi}\alpha^*-i\lambda(S_{1x}-S_{2x})\big]dt+\sqrt{\kappa}\,d\xi,\\
dS_{\ell x}&=\big[-\omega_aS_{\ell y}+\Gamma_\ell S_{\ell x}S_{\ell z}\big]dt+\sqrt{\Gamma_\ell}\,S_{\ell z}\,dW_{\ell x},\\
dS_{\ell y}&=\big[\omega_aS_{\ell x}-\lambda_\ell XS_{\ell z}+\Gamma_\ell S_{\ell y}S_{\ell z}\big]dt+\sqrt{\Gamma_\ell}\,S_{\ell z}\,dW_{\ell y},\\
dS_{\ell z}&=\big[\lambda_\ell XS_{\ell y}-\Gamma_\ell(S_{\ell x}^2+S_{\ell y}^2)\big]dt-\sqrt{\Gamma_\ell}\big(S_{\ell x}dW_{\ell x}+S_{\ell y}dW_{\ell y}\big),
\end{align}
with $X=2{\rm Re}\,\alpha$, $\lambda_1=-\lambda_2=\lambda$, $\Gamma_\ell=2\gamma_\perp/N_\ell$, complex noise
$\avg{|d\xi|^2}=dt$ and real Wiener increments $dW$. The drift reproduces the mean-field equations; the cavity noise is
exact for the linear cavity dynamics, and the spin noise conserves the spin length and reproduces the coherent-state
variance $j/2$ at $\ket{\downarrow}$. Initial values: $\alpha$ complex Gaussian with $\avg{|\alpha|^2}=1/2$,
$S_{\ell z}=-j_\ell$, and $S_{\ell x},S_{\ell y}$ Gaussian with variance $j_\ell/2$. We integrate 200 trajectories with a
Heun scheme ($dt=0.01\,\kappa^{-1}$), project the spins back onto their sphere after each step, discard a transient of
$200\,\kappa^{-1}$ and record $3000\,\kappa^{-1}$. $\avg{\hat n}=\avg{|\alpha|^2}_W-1/2$.

\textit{Observables.} $G_{AB}(\tau)$ is the symmetrized correlation averaged over trajectories and time;
$-\Gamma_{\rm lc}+i\Omega_q$ is extracted with the same harmonic inversion as in \S VI, keeping only poles that carry at
least $10\%$ of the variance of $S_{1x}$. $\mathcal C(\Omega_q)$ is the magnitude coherence of the trajectory spectra of
$S_{1x}$ and $S_{2x}$, averaged over trajectories and over the frequency bins within the peak. Error bars are standard
deviations over three noise realizations.

\textit{Validation.} Table~\ref{tab:twaval} compares the TWA with the exact results. Below threshold the quality factors agree to within
$10$--$15\%$ for $N_1=5$ and $10$, including their dependence on $g$ and on $N$, and so do the photon numbers. At $N_1=10$
above threshold the exact $Q$ increases with the cutoff, from $1.11$ ($N_{\rm cav}=30$) to $1.37$ ($N_{\rm cav}=40$),
and is still growing; the TWA value, $1.22$, lies within this range, so the comparison is limited by the cutoff. At $N_1=5$ above threshold the exact fit returns a mode at $0.41\kappa$, far from the
Hopf frequency, and the comparison is not conclusive.
Above threshold the TWA photon number ($15.7$) exceeds the exact value at every cutoff ($9.25$ at
$N_{\rm cav}=40$), consistent with the unconverged cutoff of the exact calculation (\S VI). A second-order cumulant expansion was also tested: it reproduces the
photon numbers below threshold but becomes unstable above it, because the Gaussian closure cannot represent a limit
cycle without U(1) symmetry, and it was not used.

\begin{table}[ht]
\centering
\small
\begin{tabular}{lrrrr}
\hline\hline
$\psi=-\pi/2$ & $Q$ exact & $Q$ TWA & $\avg{\hat n}$ exact & $\avg{\hat n}$ TWA\\
\hline
$N_1=10$, $g=1.0$ & 2.36 & 2.14 & 0.735 & 0.741\\
$N_1=10$, $g=1.5$ & 3.17 & 2.84 & 1.459 & 1.43\\
$N_1=10$, $g=2.0$ & 3.05 & 2.94 & 3.25 & 3.65\\
$N_1=10$, $g=3.28$ & 1.11--1.37$^*$ & 1.22 & 8.09--9.25$^*$ & 15.7\\
$N_1=5$, $g=1.5$ & 2.11 & 1.81 & 1.24 & 1.22\\
$N_1=5$, $g=3.28$ & 0.31$^*$ & 0.75 & 5.93$^*$ & 9.5\\
$N_1=15$, $g=3.28$ & 0.36$^*$ & 1.63 & 10.2$^*$ & 22.3\\
$N_1=15$, $g=1.5$ & 2.74--3.50$^\dagger$ & 3.29 & 1.40--1.64$^\dagger$ & 1.65\\
\hline\hline
\end{tabular}
\caption{TWA versus exact master equation ($N_2/N_1=0.4$). $^*$Exact values above threshold are not converged in the
cavity cutoff (range: $N_{\rm cav}=30$--$40$ at $N_1=10$; $N_{\rm cav}=30$ at $N_1=5$ and $15$). At $N_1=5$ and $15$, $g=3.28$, the exact fit returns
$\Omega_q=0.41\kappa$ and $0.53\kappa$, far from the Hopf frequency; the TWA gives $1.21\kappa$ and $1.24\kappa$. TWA entries for $g=1.0$, $2.0$ and $N_1=10$, $g=1.5$ are from independent runs; repeated runs scatter by
$\sim5\%$. $^\dagger$Below threshold at $N_1=15$ the cutoff still matters: $N_{\rm cav}=12$, 15, 20 give $\avg{\hat n}=1.40$, 1.54, 1.64 and $Q=2.74$, 3.15, 3.50.}
\label{tab:twaval}
\end{table}

\textit{Results.} Main-text Fig.~3 summarizes the TWA results; Table~\ref{tab:twascal} lists the scaling series.
Above threshold $Q$ grows without bound with a local exponent that approaches one ($0.70$ for $N_1=100\to200$, $0.83$ for
$200\to500$, $1.0$ for $500\to1000$), and $\mathcal C(\Omega_q)\to1$. Below threshold $Q$ and $\mathcal C$ saturate for
$N_1\gtrsim50$. In the $\psi$ scan at $N_1=1000$ an oscillatory mode exists only for $\psi/\pi\in[-0.667,-0.333]$
($Q=2.7$--$63$, $\mathcal C=0.88$--$1.00$), and every other point is a static superradiant state with
$\avg{\hat n}=7\times10^2$--$7\times10^4$, against $43$--$850$ inside the window. In the $g$ scan the curves for
$N_1=200$ and $1000$ coincide below threshold ($Q=1.8$, $3.4$, $16$ for $g=1$, $1.5$, $2$) and separate above it ($Q=48$
and $122$ at $g=2.3$).

\begin{table}[ht]
\centering
\small
\begin{tabular}{rrrrrr}
\hline\hline
$N_1$ & $Q$ ($g=3.28$) & $\mathcal C$ ($g=3.28$) & $\avg{\hat n}$ ($g=3.28$) & $Q$ ($g=1.5$) & $\mathcal C$ ($g=1.5$)\\
\hline
5 & $0.75\pm0.13$ & 0.15 & 9.5 & $1.81\pm0.01$ & 0.40\\
10 & $1.22\pm0.02$ & 0.30 & 15.7 & $2.84\pm0.02$ & 0.62\\
20 & $1.92\pm0.01$ & 0.44 & 28.0 & -- & --\\
50 & $2.99\pm0.01$ & 0.62 & 58.7 & $3.49\pm0.03$ & 0.81\\
100 & $4.18\pm0.04$ & 0.79 & 95.1 & -- & --\\
200 & $6.83\pm0.09$ & 0.94 & 151.5 & $3.38\pm0.03$ & 0.82\\
500 & $14.6\pm0.3$ & 0.99 & 304.5 & -- & --\\
1000 & $29.1\pm1.8$ & 0.999 & 560.4 & $3.35\pm0.03$ & 0.82\\
\hline\hline
\end{tabular}
\caption{TWA finite-size scaling at $\psi=-\pi/2$ ($N_2/N_1=0.4$).}
\label{tab:twascal}
\end{table}

\section{Effective coupling \texorpdfstring{$J_{\rm eff}(\psi)$}{Jeff(psi)}}

The static response of the cavity ($s=0$) to the spin drive follows from Eq.~\eqref{eqS:beta} with $\dot\beta=0$. This
is not an adiabatic elimination requiring $\kappa\gg\lambda\sqrt{N_\ell},\omega_a$; it is the exact $s=0$ response of
the mean-field equations. Together with its conjugate it gives
\begin{equation}
\begin{pmatrix}\kappa+i\omega_c&\kappa\mu e^{-i\psi}\\ \kappa\mu e^{i\psi}&\kappa-i\omega_c\end{pmatrix}
\begin{pmatrix}\beta\\ \beta^*\end{pmatrix}=\begin{pmatrix}-i\lambda\Delta S_x\\ +i\lambda\Delta S_x\end{pmatrix},
\end{equation}
with determinant $\kappa^2+\Omega_c^2=P$. By Cramer's rule,
\begin{equation}
X_\varphi=\beta+\beta^*=-\frac{2\lambda F(\psi)}{\kappa^2+\Omega_c^2}\Delta S_x,
\end{equation}
which defines the static cavity-mediated coupling
\begin{equation}
J_{\rm eff}(\psi)=\lambda^2\frac{F(\psi)}{\kappa^2+\Omega_c^2}.
\end{equation}
The same factor $F(\psi)$ controls the thresholds, Eqs.~\eqref{eqS:st}--\eqref{eqS:hopf}. For $\mu=1.188$ and
$\omega_c=\kappa$, $J_{\rm eff}$ vanishes at the window edges and is negative inside the window
[$J_{\rm eff}(-\pi/2)\approx-0.32\lambda^2/\kappa$] and positive outside it [$J_{\rm eff}(0)\approx+1.70\lambda^2/\kappa$].
$J_{\rm eff}$ is used only to identify the common factor $F(\psi)$; it is not a dynamical effective Hamiltonian,
since $\kappa$ is not large compared with $\lambda\sqrt{N_\ell}$ and $\omega_a$. In particular, the sign change of
$J_{\rm eff}$ does not by itself predict the $\psi$ dependence of $I(S_1{:}S_2)$, which is largest just outside the
left window edge and minimal inside the window (\S XII).

\section{Proof of the \texorpdfstring{$\mathbb Z_2$}{Z2} theorem}

The parity $\hat\Pi=\exp[i\pi(\hat a^\dagger\hat a+\hat S_{1z}+\hat S_{2z}+(N_1+N_2)/2)]$ satisfies $\hat\Pi^2=1$,
$\hat\Pi\hat a\hat\Pi^\dagger=-\hat a$ and $\hat\Pi\hat S^\pm_\ell\hat\Pi^\dagger=-\hat S^\pm_\ell$.

\textit{Theorem.} If the steady state of Eq.~\eqref{eqS:ME} is unique, $\avg{\hat O}_\infty=0$ for every odd operator
$\hat O$ ($\hat\Pi\hat O\hat\Pi^\dagger=-\hat O$).

\textit{Proof.} (1) $\hat X_\varphi$ and $\Delta\hat S_x$ are odd, so their product is even; $\hat a^{\dagger2}$ and
$\hat a^2$ are even, so $\hat H_p$ is even; hence $\hat\Pi\hat H\hat\Pi^\dagger=\hat H$. (2) Every jump operator
($\hat a$, $\hat S^-_\ell$, $\hat S_{\ell z}$) is either odd or even, so $\mathcal P[\hat L\rho\hat L^\dagger]=\hat L
\mathcal P[\rho]\hat L^\dagger$ and $\mathcal P[\hat L^\dagger\hat L\rho]=\hat L^\dagger\hat L\mathcal P[\rho]$, with
$\mathcal P[\rho]=\hat\Pi\rho\hat\Pi^\dagger$. (3) Hence $\mathcal P\circ\mathcal L=\mathcal L\circ\mathcal P$, so
$\mathcal P[\rho_\infty]$ is also a steady state and uniqueness forces $\mathcal P[\rho_\infty]=\rho_\infty$.
(4) For odd $\hat O$, $\avg{\hat O}_\infty=\mathrm{Tr}[\hat O\mathcal P[\rho_\infty]]=-\avg{\hat O}_\infty=0$.
$\square$

\textit{Corollaries.} (a) $\avg{\hat a}_\infty=0$. (b) $\avg{\hat S_{\ell x}}_\infty=\avg{\hat S_{\ell y}}_\infty=0$.
(c) $\rho_{12}=\mathrm{Tr}_{\rm cav}\rho_\infty$ is block diagonal in the eigenspaces of the two-spin parity
$\hat\Pi_1\hat\Pi_2$. This block structure does not imply a positive partial transpose. Numerically,
$|\avg{\hat a}_\infty|<10^{-10}$ and the logarithmic negativity~\cite{Vidal2002,Plenio2005} satisfies
$E_N(\rho_{12})<10^{-9}$ in every simulation. Since PPT states may be bound entangled~\cite{Horodecki1998}, we make no
separability claim; none of the witnesses relies on entanglement.

\section{Numerical methods and convergence}

\textit{Steady states.} Hilbert space: cavity ($N_{\rm cav}=30$) $\otimes$ spin 1 (dimension 11) $\otimes$ spin 2
(dimension 5), total dimension 1650. The Liouvillian has a single zero eigenvalue at all reported parameters, so the
steady state is unique and independent of the initial state, which affects only the transient.

\textit{Liouvillian.} All steady states (main-text Fig.~2(b),(c),(e),(f), Figs.~\ref{fig:nphoton} and
\ref{fig:S1}--\ref{fig:S3}, and the single-ensemble control, Fig.~\ref{fig:single})
and all two-time results use the same Liouvillian, Eq.~\eqref{eqS:ME}, with
$\Gamma_\ell=(2\gamma_\perp-\gamma_\phi)/N_\ell$ so that both ensembles have the transverse damping $\gamma_\perp$
($\gamma_\phi=0$ unless stated; for $N_2=0$ the second ensemble is absent). It is assembled in a banded representation
and checked against an explicit Kronecker-product construction (maximum relative difference $3\times10^{-16}$); its
spectrum satisfies ${\rm Re}\,\lambda\le0$ to $5\times10^{-14}$. Steady states are obtained by time integration,
continued from the previous point of each parameter scan, until $\|\mathcal L\rho\|/\|\rho\|<5\times10^{-4}$ or
$t=150\,\kappa^{-1}$; over all steady states of main-text Fig.~2 and Figs.~\ref{fig:nphoton}--\ref{fig:S3} and
\ref{fig:single} the residual is at most $3.5\times10^{-3}$ (median $4\times10^{-4}$; the thermal-photon run of
Table~\ref{tab:exactother}, $1.4\times10^{-2}$, is the only exception);
it exceeds $10^{-3}$ only at $g=3.28$ for $0\le\psi/\pi\le0.17$, at $\psi=0$ for $g\ge3.28$ and at $\psi=-\pi/2$ for $g\ge4.5$, where $\avg{\hat n}\gtrsim11$ and the cutoff is most strained; no conclusion rests on these points.

\textit{Entropies.} Partial traces are taken from $\rho_\infty$; $S(\rho)=-\sum_ip_i\ln p_i$ (eigenvalues below
$10^{-12}$ discarded) and $I(A{:}B)=S(\rho_A)+S(\rho_B)-S(\rho_{AB})$, in natural-log units. The Wehrl entropy
$S_W=-\int Q\ln Q\,d^2\alpha$ is computed on a $200\times200$ grid with $|\alpha|\le8$.

\textit{Discord.} $J(S_1{:}S_2)$ is optimized over local von Neumann measurements on $S_2$ (dimension 5): 300
Haar-random samples followed by Nelder--Mead refinement of the best three (25 real parameters), converged to $\sim1\%$.
The protocol is repeated measuring $S_1$ (dimension 11, 121 parameters). Restarts from the best five candidates at
$\psi/\pi=-0.5$, 0, $+0.5$ spread by less than 1.5\%. The asymmetry $\Delta D=D({\rm meas}\,S_1)-D({\rm meas}\,S_2)>0$
exceeds the optimizer noise by a factor 5--10 where it is resolved ($g\gtrsim0.8$).

\textit{Convergence.} Below threshold the cutoff $N_{\rm cav}=30$ is ample (photon numbers $\lesssim3.3$, in agreement
with the TWA); above threshold $\avg{\hat n}$ and $Q$ are not converged at $N_{\rm cav}=40$ and are lower bounds
(\S VI); $t_{\max}=30\to60\,\kappa^{-1}$:
$<10^{-3}$ change; discord sample budget $100\to1000$: $<2\%$ change. The two-time method and its validation are
described in \S VI.
Figure~\ref{fig:ncav} follows the steady-state observables at $g=3.28$ for $N_{\rm cav}=20$--$40$ at three phases,
starting from the empty cavity:
\begin{itemize}
\item \textit{Window edge} ($\psi/\pi=-0.667$): all quantities converge; from $N_{\rm cav}=30$ to $40$, $\avg{\hat n}$
changes by $2.6\%$ ($4.58\to4.70$), $I(S_1{:}S_2)$ by $4\%$ and $D/I$ by less than $1\%$.
\item \textit{Window center} ($\psi=-\pi/2$): $D/I$ is stable to within $5\%$ ($0.423$, $0.417$, $0.405$ measuring
$S_2$ and $0.609$, $0.590$, $0.582$ measuring $S_1$ for $N_{\rm cav}=30$, $35$, $40$), whereas $I(S_1{:}S_2)$ and $D$
separately decrease with the cutoff ($I=0.269$, $0.207$, $0.162$). In the limit-cycle regime the one-time correlations
are therefore used only through $D/I$; the minimum of $I(S_1{:}S_2)$ inside the window (\S XII) deepens with the cutoff.
\item \textit{Static superradiant phase} ($\psi=0$): $\avg{\hat n}$ grows almost linearly with the cutoff ($8.75$,
$11.3$, $13.9$, $16.6$, $19.3$ for $N_{\rm cav}=20$--$40$; mean field: $\sim880$), and $D/I$ decreases ($0.296$, $0.264$, $0.224$, $0.183$, $0.155$ measuring $S_2$; measuring $S_1$ it levels off at $0.42$, still below the window value $0.58$). The values of $D/I$ quoted for the static superradiant phases (\S XI) are therefore
upper bounds, and the contrast with the window ($0.07$, $0.15$, $0.20$, $0.23$, $0.25$ for $N_{\rm cav}=20$--$40$) is a lower
bound. The photon-number dip $1-\avg{\hat n}(-0.667\pi)/\avg{\hat n}(0)$ likewise grows: $49\%$, $60\%$, $67\%$, $72\%$, $76\%$. At $N_{\rm cav}=40$ the steady-state residual at $\psi=0$ is $2\times10^{-3}$, the largest of these runs.
\end{itemize}
\textit{Initial-state independence.} At $N_{\rm cav}=30$ the steady states reached from the empty cavity coincide with
those of the production scans, which start from the neighboring parameter point: $\avg{\hat n}$ agrees to
$3\times10^{-5}$ ($\psi=0$), $5\times10^{-4}$ ($\psi=-\pi/2$) and $1.2\times10^{-3}$ ($\psi=-0.667\pi$), and
$I(S_1{:}S_2)$ and $D/I$ to $1\%$, consistent with the steady-state residual.

\section{Quantum fraction of the inter-ensemble correlations}

The steady states are computed with the model of Eq.~\eqref{eqS:ME}. The ratio $D/I$ depends on the phase, on the
coupling and on which ensemble is measured, and it is this dependence that carries information:
\begin{itemize}
\item \textit{Inside the window} ($g=3.28$), $D/I\simeq0.42$--$0.44$ measuring $S_2$ and $0.59$--$0.62$ measuring
$S_1$ for $-0.67\le\psi/\pi\le-0.50$, decreasing towards the right edge to $0.29$ ($0.54$) at $\psi/\pi=-0.33$
[main-text Fig.~2(f)].
\item \textit{Outside the window} the correlations are mostly classical: $D/I=0.01$--$0.07$
($0.32$--$0.35$) for $\psi/\pi\le-0.83$, $0.12$ ($0.41$) at the maximum of
$I(S_1{:}S_2)$, $\psi/\pi=-0.75$, $0.24$--$0.27$ ($0.47$--$0.54$) for $-0.25\le\psi/\pi<0$ and $0.12$--$0.29$
($0.36$--$0.52$) for $\psi>0$ (values in parentheses: measurement on $S_1$). These values decrease with the cutoff and are upper bounds (\S X).
\item \textit{Versus coupling} at $\psi=-\pi/2$, $D/I$ decreases from $0.89$ at $g=0.25$, where the correlations
are weak but mostly quantum, to $0.42$--$0.43$ for $2.5\le g\le3.5$, and further to $0.32$ at $g=4$ and $0.19$ at
$g=5$ ($0.58$ and $0.51$ measuring $S_1$); at $\psi=0$ it falls to $0.27$--$0.28$ for $0.9\le g\le2.15$, once the
static threshold $g_{\rm st}\simeq0.68$ is crossed, and decreases slowly to $0.17$ ($0.40$ measuring $S_1$) at $g=5$ (Fig.~\ref{fig:S3}).
The points at $g\ge4.5$ have the largest steady-state residuals (\S X).
\end{itemize}
At equal coupling ($g=3.28$) the quantum share in the Hopf window is therefore larger than in either superradiant
phase. This is expected: in a static superradiant phase the parity-symmetric steady state is close to an equal mixture
of the two symmetry-broken branches, in which both ensembles share the same branch, a correlation that is classical;
the parity block structure of $\rho_{12}$
(corollary (c) of \S IX) is compatible with these values but does not fix them. The uncertainty of $D$ combines
optimizer convergence ($\sim1\%$) and the sample-budget dependence ($\sim2\%$, \S X).

\section{Cavity--spin mutual information}

\begin{figure}[ht]
\centering
\figfile{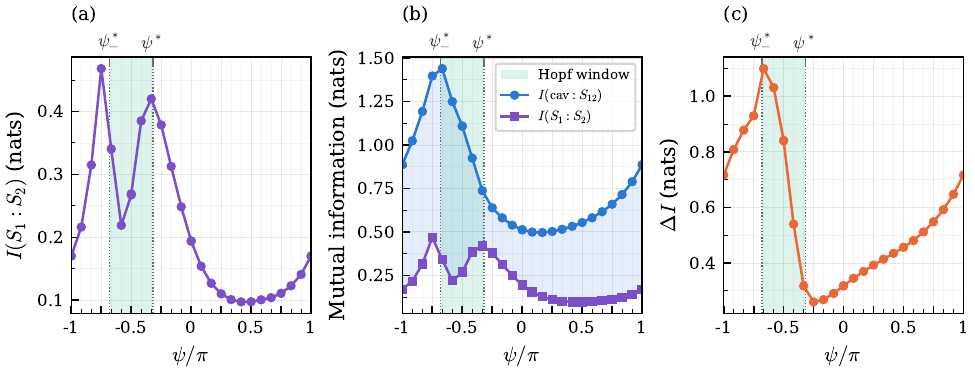}{\textwidth}
\caption{Spin--spin and cavity--spin mutual information versus $\psi$ at $g=3.28$, $r=0.8$, $\gamma_\perp=0.3\kappa$,
$N_1=10$, $N_2=4$. Green band and dotted lines: analytical window, edges $\psi^\ast_-$ and $\psi^\ast$. (a) $I(S_1{:}S_2)$.
(b) $I({\rm cav}{:}S_{12})$ (blue) and $I(S_1{:}S_2)$ (purple); shading: gap $\Delta I$. (c)
$\Delta I=I({\rm cav}{:}S_{12})-I(S_1{:}S_2)$; maximum $\Delta I_{\max}=1.10$ at $\psi/\pi=-0.67$.
Natural-log units.}
\label{fig:S1}
\end{figure}

\begin{figure}[ht]
\centering
\figfile{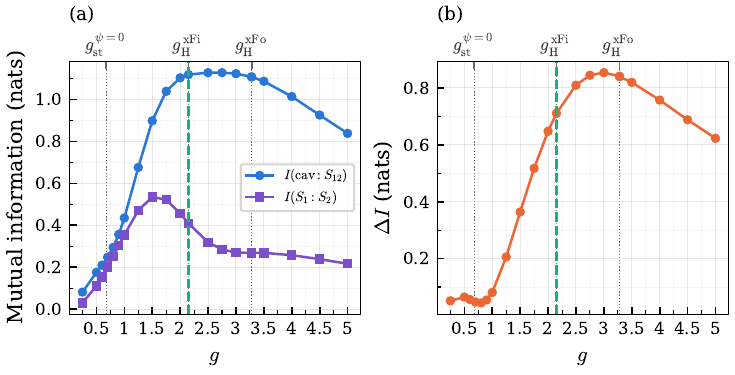}{0.75\textwidth}
\caption{Mutual information versus $g$ at $\psi=-\pi/2$ ($r=0.8$, $\gamma_\perp=0.3\kappa$, $N_1=10$, $N_2=4$). Green
dashed: $g^{\rm xFi}_{\rm Hopf}=2.15$. (a) $I({\rm cav}{:}S_{12})$ (blue) and $I(S_1{:}S_2)$ (purple). (b)
$\Delta I=I({\rm cav}{:}S_{12})-I(S_1{:}S_2)$; maximum $\approx0.85$ near $g\approx3.0$.}
\label{fig:S2}
\end{figure}

\begin{figure}[ht]
\centering
\figfile{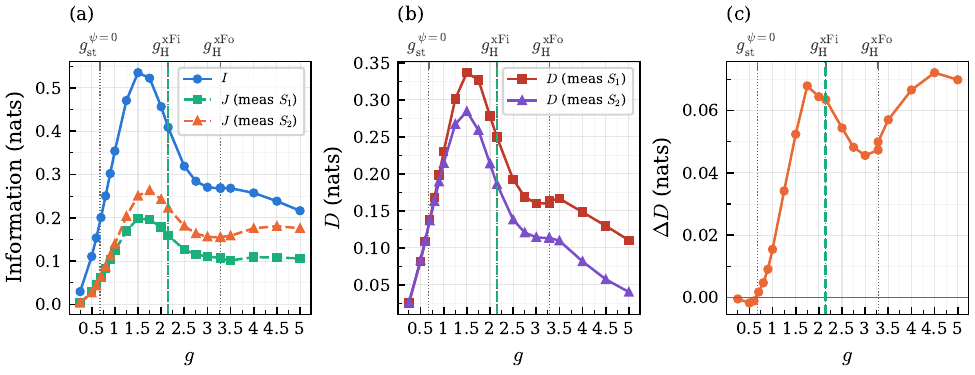}{\textwidth}
\caption{Quantum correlations versus $g$ at $\psi=-\pi/2$ (parameters as in Fig.~\ref{fig:S2}). Green dashed:
$g^{\rm xFi}_{\rm Hopf}=2.15$. (a) $I$ and $J$ for both measurement directions; $J>D$ measuring $S_2$ above $g\approx1.75$.
(b) $D$ for both measurement directions; maximum near $g\approx1.5$, below $g^{\rm xFi}_{\rm Hopf}$, with
$D({\rm meas}\,S_1)\ge D({\rm meas}\,S_2)$. (c) $\Delta D=D({\rm meas}\,S_1)-D({\rm meas}\,S_2)\ge0$ within the optimizer noise, reflecting
$N_1>N_2$.}
\label{fig:S3}
\end{figure}

Figure~\ref{fig:S1} shows the mutual-information portrait versus $\psi$ at $g=3.28$. The cavity--spin mutual
information exceeds $I(S_1{:}S_2)$ by a factor 1.7--5.9, reflecting the cavity-mediated origin of the
correlations. $I(S_1{:}S_2)$ is largest just outside the left edge of the window ($\psi/\pi=-0.75$) and at its
right edge, with a minimum inside; as discussed in \S XI, the window is distinguished by the quantum fraction $D/I$, not
by the amount of correlation. The gap $\Delta I=I({\rm cav}{:}S_{12})-I(S_1{:}S_2)$ peaks at
$\Delta I_{\max}=1.10$ at the window edge, $\psi/\pi=-0.67$. Figure~\ref{fig:S2} shows the coupling
dependence at $\psi=-\pi/2$: $I({\rm cav}{:}S_{12})$ grows up to the Hopf threshold, stays near its maximum
($\approx1.13$) between $g^{\rm xFi}_{\rm Hopf}$ and $g^{\rm xFo}_{\rm Hopf}$ and decreases beyond ($0.84$ at $g=5$),
and $\Delta I$ peaks near $g\approx3.0$.
Figure~\ref{fig:S3} shows the $I=J+D$ decomposition and the discord asymmetry
versus $g$. $I$ and $D$ reach their maxima near $g\approx1.5$, below $g^{\rm xFi}_{\rm Hopf}=2.15$, and then
decrease; measuring $S_2$, $J>D$ above $g\approx1.75$, whereas measuring $S_1$ the discord exceeds $J$ over the
whole range; $\Delta D\ge0$ within the optimizer noise, and clearly positive for $g\gtrsim0.8$. These one-time correlations characterize $\rho_\infty$ but do not mark the onset of
synchronization.

\section{Mean-field synchronization figures}

\begin{figure}[ht]
\centering
\figfile{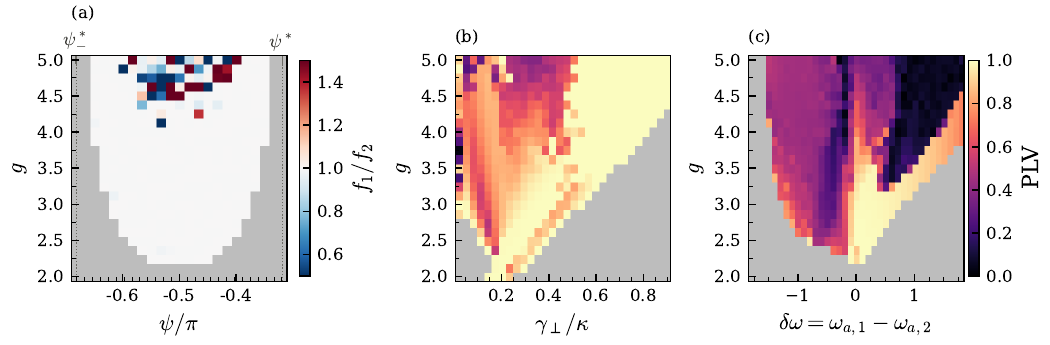}{\textwidth}
\caption{Mean-field synchronization maps ($N_2/N_1=0.4$, collective spin decay). (a) Frequency ratio $f_1/f_2$
over the window in the $(\psi/\pi,g)$ plane at $\gamma_\perp=0.3\kappa$ (gray: non-oscillating). (b) Phase-locking value
${\rm PLV}=|\avg{e^{i(\phi_1-\phi_2)}}|$ in the $(\gamma_\perp,g)$ plane at $\psi=-\pi/2$. (c) Arnold tongue:
PLV in the $(\delta\omega,g)$ plane at $\psi=-\pi/2$, $\gamma_\perp=0.3\kappa$, with $\delta\omega=\omega_{a,1}-\omega_{a,2}$.
$\phi_2$ is measured for $-\hat S_2$, which couples with the opposite sign.}
\label{fig:S4}
\end{figure}

The maps were obtained by integrating the mean-field equations (the deterministic part of the TWA equations of
\S VII, with collective decay) for $1.5\times10^3\,\kappa^{-1}$ and analysing the last $500\,\kappa^{-1}$. Figure~\ref{fig:S4}(a) shows
$f_1/f_2=1$ throughout the window above threshold, except for scattered points at $g\gtrsim4$ where the dynamics is
irregular. In the regular limit-cycle region the locking is strict, ${\rm PLV}\simeq1$, with an offset
$\phi_1-\phi_2=0.33\pi$, $0.40\pi$, $0.44\pi$ at $g=2.5$, $2.75$, $3.0$ ($\psi=-\pi/2$); since the offset is not zero, the
Kuramoto parameter $|\tfrac12\sum_\ell e^{i\phi_\ell}|=\cos[(\phi_1-\phi_2)/2]$ is $0.77$--$0.87$ although the ensembles are locked.

Above
$g\approx g^{\rm xFo}_{\rm Hopf}$ the oscillation becomes irregular and the PLV decreases (0.87, 0.75, 0.44 at
$g=3.5$, $4$, $5$). Panel (b) shows that oscillation persists up to $\gamma_\perp\approx0.5\kappa$ at $g=3$ and to larger
$\gamma_\perp$ at larger $g$. Panel (c) shows the Arnold tongue: the frequency-locked range of detuning widens with $g$,
from $|\delta\omega|\approx0.2\kappa$ at $g=2.25$ to $\approx3\kappa$ at $g\approx4$, as expected for Hopf-mediated
synchronization~\cite{Pikovsky2003}, while strict phase locking (PLV$\simeq1$) occupies a narrower band, asymmetric in
$\delta\omega$ because $N_1\ne N_2$.

\begin{figure}[ht]
\centering
\figfile{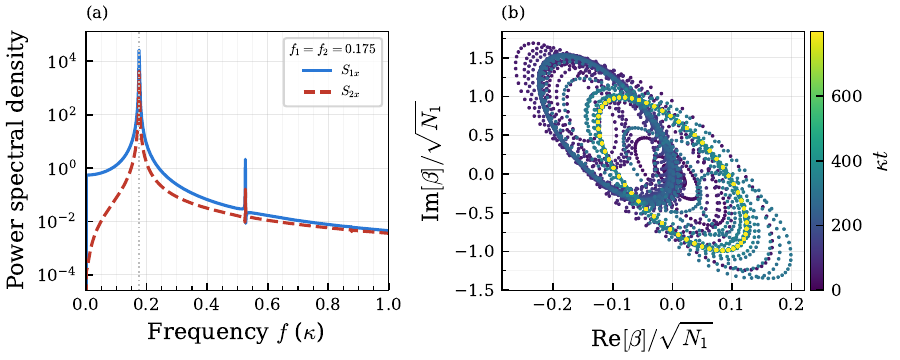}{0.85\textwidth}
\caption{Mean-field synchronization portrait at $g=2.80$, $\psi=-\pi/2$, $\gamma_\perp=0.3\kappa$ ($N_2/N_1=0.4$). (a) Power
spectra of $S_{1x}$ (solid) and $S_{2x}$ (dashed) on the limit cycle, both peaking at $f_1=f_2=0.175\kappa$. (b) Cavity
trajectory in the $({\rm Re}\,\beta,{\rm Im}\,\beta)$ plane from $t=0$, converging to the limit cycle. These are mean-field spectra; the
quantum two-time spectra are discussed in \S VI.}
\label{fig:S5}
\end{figure}

Figure~\ref{fig:S5} shows a mean-field portrait at $(g,\psi,\gamma_\perp)=(2.80,-\pi/2,0.3)$. Both spectra peak at
$f_1=f_2=0.175\kappa$; the frequency at the Hopf threshold is $\Omega/2\pi=0.157\kappa$ [Eq.~\eqref{eqS:Om}], and the shift
reflects the finite amplitude of the cycle above threshold. The phase difference is locked (PLV$=0.996$) at an offset of
$0.41\pi$ (Kuramoto parameter $0.80$). After a transient of a few hundred $\kappa^{-1}$ the cavity trajectory converges to a closed limit cycle.

\section{Parameter ranges, coupling signs and robustness}

\begin{figure}[ht]
\centering
\figfile{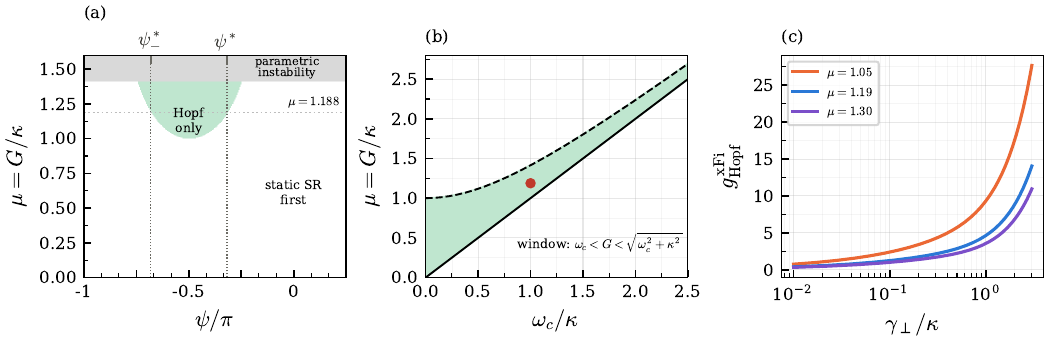}{\textwidth}
\caption{Parameter ranges (analytical, $N_2/N_1=0.4$). (a) Instability type of the normal state in the
$(\psi,\mu)$ plane at $\omega_c=\kappa$: Hopf only inside the green region; static SR first elsewhere; the cavity mode
is unstable above $\mu=\sqrt{1+\omega_c^2/\kappa^2}$. The dotted line marks $\mu=1.188$ ($r=0.8$). (b) Existence
range $\omega_c<\kappa\mu<\sqrt{\omega_c^2+\kappa^2}$ versus cavity detuning; dot: parameters used. (c) Hopf threshold
at $\psi=-\pi/2$ versus $\gamma_\perp$ for three values of $\mu$; $g_{\rm Hopf}\propto\sqrt{\gamma_\perp}$ at small
$\gamma_\perp$. Dotted lines in (a): window edges $\psi^\ast_-$ and $\psi^\ast$ at $\mu=1.188$.}
\label{fig:S6}
\end{figure}

\textit{Parameter ranges.} Figure~\ref{fig:S6} illustrates \S III: the instability type in the $(\psi,\mu)$ plane, the
existence range Eq.~\eqref{eqS:range} versus detuning, and the Hopf threshold versus $\gamma_\perp$.

\begin{figure}[ht]
\centering
\figfile{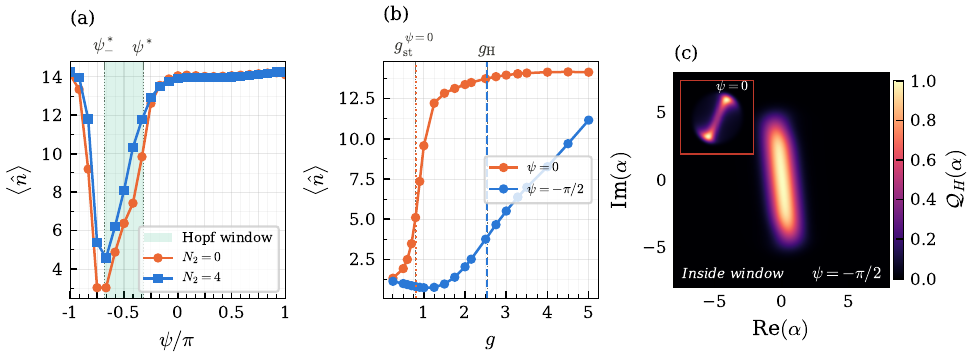}{\textwidth}
\caption{Single-ensemble control ($N_2=0$, $N_1=10$, $N_{\rm cav}=30$). (a) $\avg{\hat n}$ versus $\psi$ at
$g=3.28$ for $N_2=0$ and, for comparison, $N_2=4$; green band and dotted lines: window, edges $\psi^\ast_-$ and
$\psi^\ast$. (b) $\avg{\hat n}$ versus $g$ at $\psi=0$ and $\psi=-\pi/2$ for $N_2=0$; dotted and dashed lines: static
threshold at $\psi=0$ ($g_{\rm st}^{\psi=0}=0.80$) and Hopf threshold ($g_{\rm H}=2.54$) for $N_2=0$. (c) Husimi function $\mathcal Q_H(\alpha)$ at $\psi=-\pi/2$ for $N_2=0$ (inset: $\psi=0$).}
\label{fig:single}
\end{figure}

\textit{Single ensemble ($N_2=0$).} Eq.~\eqref{eqS:general} reduces to $D_2(s)[K(s)D_1(s)-F(\psi)\omega_1\lambda_1^2N_1]$,
so the window and the Hopf-only instability persist for one ensemble. The photonic observables, controlled by $F(\psi)$,
certify the oscillatory regime; synchronization is undefined for one ensemble. A master-equation run at $N_2=0$
(dimension 330, Fig.~\ref{fig:single}) confirms it: at $g=3.28$, $\avg{\hat n}$ drops by 79\% and $S_W$ by
23\% inside the window, with the minimum at $\psi/\pi=-0.67$ (panel (a)). Since $N_{\rm eff}=N_1$ instead of $N_1+N_2$, the thresholds in $g=\sqrt{N_1}\lambda/\kappa$ rise by
$\sqrt{(N_1+N_2)/N_1}=\sqrt{1.4}$: $g_{\rm st}^{\psi=0}=0.80$ and $g_{\rm Hopf}=2.54$ (panel (b)). The Husimi function $\mathcal Q_H(\alpha)$ (panel (c)) is a single
stripe at $\psi=-\pi/2$ for $N_2=0$ (inset: $\psi=0$). It is nearly identical to main-text Fig.~2(c) ($N_2=4$), since the cavity state is controlled by $F(\psi)$ alone.

\textit{Coupling signs and magnitudes.} The rotation $e^{i\pi\hat S_{2z}}$ maps $\lambda_2\to-\lambda_2$ and
$\hat S^-_2\to-\hat S^-_2$, leaving Eq.~\eqref{eqS:ME} invariant. Hence all observables are identical for
$\lambda_2=\pm\lambda_1$, and only the sign of $\avg{\hat S^+_1\hat S^-_2}$ changes. For $|\lambda_2|\ne|\lambda_1|$ the
window and the instability type are unchanged, since Eq.~\eqref{eqS:general} depends on the couplings only through
$\lambda_\ell^2$.

\textit{Robustness to additional noise.} Additional dephasing and inhomogeneous damping only renormalize
$\gamma_{\perp,\ell}$, and unequal frequencies enter through $\omega_\ell$, so the window and the instability type are
unchanged. Thermal photons do not enter Eq.~\eqref{eqS:beta}. In mean field, the oscillatory region persists up to
$\gamma_\perp\approx0.5\kappa$ at $g=3$ and over an inter-ensemble detuning range that widens with $g$; at $g\approx3$ it
spans pump-phase deviations of about $\pm0.1\pi$ around $\psi=-\pi/2$. Full master-equation runs with
$n_{\rm th}=0.1$ and with $\gamma_\phi=0.1\kappa$ give $Q=0.99$ and $1.13$, against $1.11$ without them
(Table~\ref{tab:exactother}).

\section{Robustness of the emergence and the locked phase difference}

\begin{figure}[ht]
\centering
\figfile{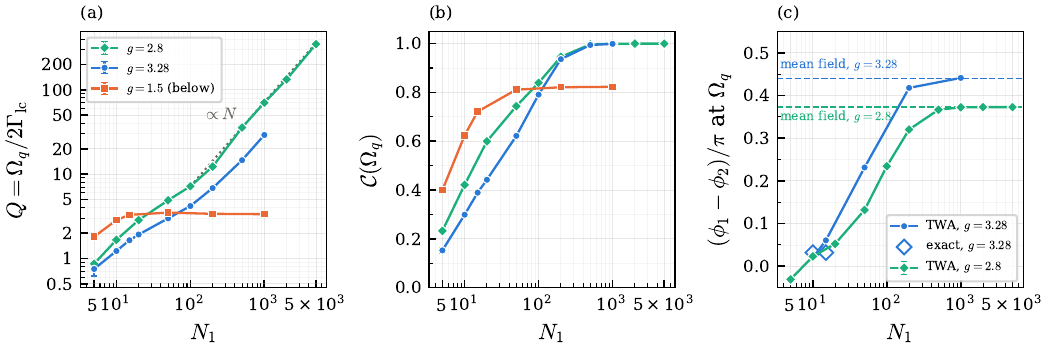}{\textwidth}
\caption{TWA finite-size scaling at $\psi=-\pi/2$ ($N_2/N_1=0.4$, $\gamma_\perp=0.3\kappa$; three noise realizations per
point). (a) $Q$ and (b) $\mathcal C(\Omega_q)$ at $g=2.8$, inside the regular mean-field limit cycle, compared with
$g=3.28$ and $g=1.5$ (Fig.~3 of the main text). (c) Locked phase difference between the ensembles at $\Omega_q$,
$\phi_1-\phi_2=\arg\avg{\tilde S_{1x}(\Omega_q)[-\tilde S_{2x}(\Omega_q)]^*}$ (first harmonic of $S_{\ell x}$, staggered
frame); dashed lines: the same quantity on the mean-field limit cycle; open diamonds: exact results for $N_1=10$ and $15$,
$N_{\rm cav}=30$, $\arg[-P_{12}(\omega)]$ at the frequency of the Hopf mode ($1.19\kappa$ and $1.24\kappa$; at $N_1=15$
the fitted $\Omega_q=0.53\kappa$ is set by the cutoff, \S VI). The $g=2.8$ series is extended to $N_1=2000$ and $5000$ (one realization).}
\label{fig:S10}
\end{figure}

The value $g=3.28$ used in main-text Fig.~3 lies at $g^{\rm xFo}_{\rm Hopf}$, the upper edge of the regular mean-field
limit cycle (\S XIII). To show that the emergence does not depend on this choice we repeated the scaling series at
$g=2.8$, well inside the regular limit cycle (Fig.~\ref{fig:S10}). Above threshold $Q$ again grows without bound, from
$1.66\pm0.01$ at $N_1=10$ to $70\pm4$ at $N_1=1000$, $133$ at $2000$ and $348$ at $5000$ (local exponents $0.98$, $0.92$,
$1.05$), with larger values than at $g=3.28$ ($29\pm2$ at $N_1=1000$), and $\mathcal C(\Omega_q)\to1$.

The cross spectrum also gives the relative phase of the two ensembles at the oscillation frequency, i.e.\ the phase
difference of the first harmonics of $S_{1x}$ and $-S_{2x}$. It grows with system size, from $\approx0$ for $N_1\le10$ to
$0.372\pi$ at $N_1=1000$ and $0.3727\pi$ at $N_1=2000$ and $5000$ for $g=2.8$, and to $0.441\pi$ at $N_1=1000$ for
$g=3.28$. The same quantity evaluated on the mean-field limit cycle is $0.373\pi$ and $0.440\pi$: the TWA converges to
it. (The instantaneous phase difference of \S XIII, $0.41\pi$ and $0.48\pi$, differs slightly because the spin orbits
are not circular.) At $N_1=10$, $g=3.28$ the TWA value, $0.020\pi$, agrees with the exact one, $0.03\pi$. For small ensembles quantum noise
washes out the relative phase; as $N$ grows the slow quantum mode acquires both the coherence and the locked phase
difference of the mean-field limit cycle. This is a further, independent signature that the mode whose quality
factor grows with $N$ is the synchronized limit cycle.

\begin{figure}[ht]
\centering
\figfile{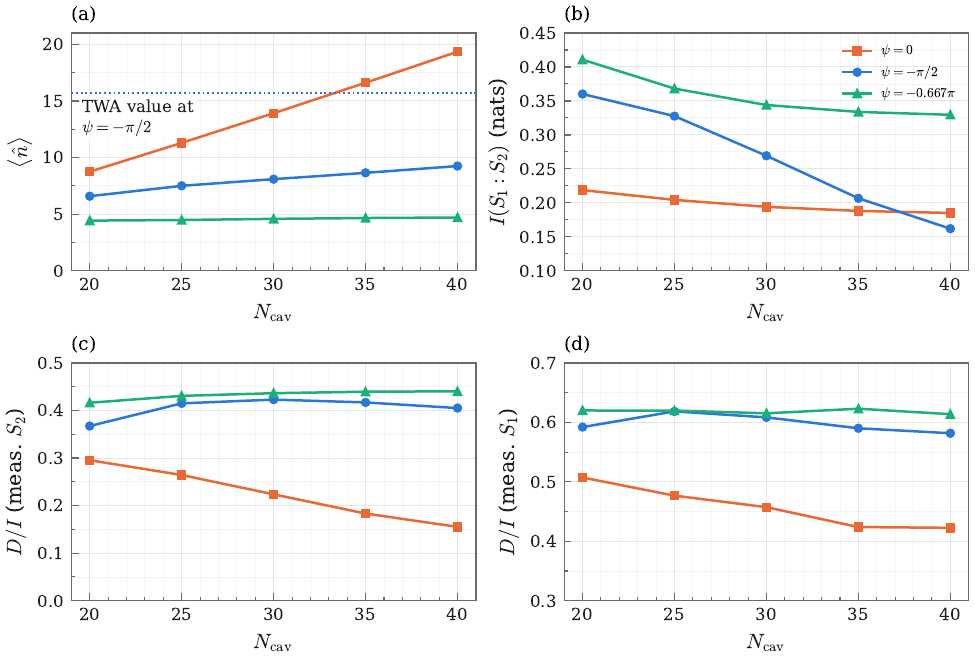}{\textwidth}
\caption{Cavity-cutoff convergence of the steady-state observables at $g=3.28$ ($N_1=10$, $N_2=4$, empty-cavity
initial state; \S X). (a) $\avg{\hat n}$; dotted line: TWA value at $\psi=-\pi/2$ ($15.7$); at $\psi=0$ the mean-field value ($\sim880$) is off scale. (b) $I(S_1{:}S_2)$. (c) $D/I$
measuring $S_2$ and (d) measuring $S_1$. Squares: $\psi=0$ (static superradiance); circles: $\psi=-\pi/2$; triangles:
$\psi=-0.667\pi$ (window edge).}
\label{fig:ncav}
\end{figure}


\begin{thebibliography}{0}%
\makeatletter
\providecommand \@ifxundefined [1]{%
 \@ifx{#1\undefined}
}%
\providecommand \@ifnum [1]{%
 \ifnum #1\expandafter \@firstoftwo
 \else \expandafter \@secondoftwo
 \fi
}%
\providecommand \@ifx [1]{%
 \ifx #1\expandafter \@firstoftwo
 \else \expandafter \@secondoftwo
 \fi
}%
\providecommand \natexlab [1]{#1}%
\providecommand \enquote  [1]{``#1''}%
\providecommand \bibnamefont  [1]{#1}%
\providecommand \bibfnamefont [1]{#1}%
\providecommand \citenamefont [1]{#1}%
\providecommand \href@noop [0]{\@secondoftwo}%
\providecommand \href [0]{\begingroup \@sanitize@url \@href}%
\providecommand \@href[1]{\@@startlink{#1}\@@href}%
\providecommand \@@href[1]{\endgroup#1\@@endlink}%
\providecommand \@sanitize@url [0]{\catcode `\\12\catcode `\$12\catcode `\&12\catcode `\#12\catcode `\^12\catcode `\_12\catcode `\%12\relax}%
\providecommand \@@startlink[1]{}%
\providecommand \@@endlink[0]{}%
\providecommand \url  [0]{\begingroup\@sanitize@url \@url }%
\providecommand \@url [1]{\endgroup\@href {#1}{\urlprefix }}%
\providecommand \urlprefix  [0]{URL }%
\providecommand \Eprint [0]{\href }%
\providecommand \doibase [0]{https://doi.org/}%
\providecommand \selectlanguage [0]{\@gobble}%
\providecommand \bibinfo  [0]{\@secondoftwo}%
\providecommand \bibfield  [0]{\@secondoftwo}%
\providecommand \translation [1]{[#1]}%
\providecommand \BibitemOpen [0]{}%
\providecommand \bibitemStop [0]{}%
\providecommand \bibitemNoStop [0]{.\EOS\space}%
\providecommand \EOS [0]{\spacefactor3000\relax}%
\providecommand \BibitemShut  [1]{\csname bibitem#1\endcsname}%
\let\auto@bib@innerbib\@empty
%</preamble>
\end{thebibliography}%


\begin{thebibliography}{99}
\bibitem{Hepp1973} K. Hepp and E. H. Lieb, Ann. Phys. (N.Y.) \textbf{76}, 360 (1973).
\bibitem{Wang1973} Y. K. Wang and F. T. Hioe, Phys. Rev. A \textbf{7}, 831 (1973).
\bibitem{Garraway2011} B. M. Garraway, Phil. Trans. R. Soc. A \textbf{369}, 1137 (2011).
\bibitem{Emary2003} C. Emary and T. Brandes, Phys. Rev. E \textbf{67}, 066203 (2003).
\bibitem{Larson2021} J. Larson and T. Mavrogordatos, \textit{The Jaynes--Cummings Model and Its Descendants} (IOP Publishing, 2021).
\bibitem{Alvarez2026} J. D. \'Alvarez-Cuartas and J. H. Reina, Phys. Rev. Res. \textbf{8}, 033035 (2026).
\bibitem{Ritsch2013} H. Ritsch, P. Domokos, F. Brennecke, and T. Esslinger, Rev. Mod. Phys. \textbf{85}, 553 (2013).
\bibitem{Mivehvar2021} F. Mivehvar, F. Piazza, T. Donner, and H. Ritsch, Adv. Phys. \textbf{70}, 1 (2021).
\bibitem{Kirton2019} P. Kirton, M. M. Roses, J. Keeling, and E. G. Dalla Torre, Adv. Quantum Technol. \textbf{2}, 1800043 (2019).
\bibitem{Baumann2010} K. Baumann, C. Guerlin, F. Brennecke, and T. Esslinger, Nature \textbf{464}, 1301 (2010).
\bibitem{Klinder2015} J. Klinder \textit{et al.}, Proc. Natl. Acad. Sci. U.S.A. \textbf{112}, 3290 (2015).
\bibitem{Dogra2019} N. Dogra \textit{et al.}, Science \textbf{366}, 1496 (2019).
\bibitem{Baksic2014} A. Baksic and C. Ciuti, Phys. Rev. Lett. \textbf{112}, 173601 (2014).
\bibitem{SafaviNaini2018} A. Safavi-Naini \textit{et al.}, Phys. Rev. Lett. \textbf{121}, 040503 (2018).
\bibitem{Mivehvar2024} F. Mivehvar, Phys. Rev. Lett. \textbf{132}, 073602 (2024).
\bibitem{Li2025} Y. Li \textit{et al.}, Sci. Adv. \textbf{11}, eady5649 (2025).
\bibitem{Vaidya2024} G. M. Vaidya \textit{et al.}, Phys. Rev. A \textbf{109}, 033718 (2024).
\bibitem{Parra2020} A. Parra-L\'opez and J. Bergli, Phys. Rev. A \textbf{101}, 062104 (2020).
\bibitem{Xu2014} M. Xu, D. A. Tieri, E. C. Fine, J. K. Thompson, and M. J. Holland, Phys. Rev. Lett. \textbf{113}, 154101 (2014).
\bibitem{Sonar2018} S. Sonar \textit{et al.}, Phys. Rev. Lett. \textbf{120}, 163601 (2018).
\bibitem{Murtadho2023} T. Murtadho, S. Vinjanampathy, and J. Thingna, Phys. Rev. Lett. \textbf{131}, 030401 (2023).
\bibitem{Tindall2020} J. Tindall, C. S\'anchez Mu\~noz, B. Bu\v{c}a, and D. Jaksch, New J. Phys. \textbf{22}, 013026 (2020).
\bibitem{Roulet2018} A. Roulet and C. Bruder, Phys. Rev. Lett. \textbf{121}, 063601 (2018).
\bibitem{Keeling2010} J. Keeling, M. J. Bhaseen, and B. D. Simons, Phys. Rev. Lett. \textbf{105}, 043001 (2010).
\bibitem{Bhaseen2012} M. J. Bhaseen, J. Mayoh, B. D. Simons, and J. Keeling, Phys. Rev. A \textbf{85}, 013817 (2012).
\bibitem{Walter2014} S. Walter, A. Nunnenkamp, and C. Bruder, Phys. Rev. Lett. \textbf{112}, 094102 (2014).
\bibitem{Lorch2016} N. L\"orch, E. Amitai, A. Nunnenkamp, and C. Bruder, Phys. Rev. Lett. \textbf{117}, 073601 (2016).
\bibitem{Mari2013} A. Mari, A. Farace, N. Didier, V. Giovannetti, and R. Fazio, Phys. Rev. Lett. \textbf{111}, 103605 (2013).
\bibitem{Ameri2015} V. Ameri \textit{et al.}, Phys. Rev. A \textbf{91}, 012301 (2015).
\bibitem{Bastidas2010} V. M. Bastidas, J. H. Reina, C. Emary, and T. Brandes, Phys. Rev. A \textbf{81}, 012316 (2010).
\bibitem{Bastidas2015} V. M. Bastidas \textit{et al.}, Phys. Rev. A \textbf{91}, 053824 (2015).
\bibitem{Laskar2020} A. W. Laskar \textit{et al.}, Phys. Rev. Lett. \textbf{125}, 013601 (2020).
\bibitem{Krithika2022} V. R. Krithika, P. Solanki, S. Vinjanampathy, and T. S. Mahesh, Phys. Rev. A \textbf{105}, 062206 (2022).
\bibitem{Nigg2018} S. E. Nigg, Phys. Rev. A \textbf{97}, 013811 (2018).
\bibitem{Hassan1982} S. S. Hassan, G. P. Hildred, R. R. Puri, and S. V. Lawande, J. Phys. B \textbf{15}, 1029 (1982).
\bibitem{Agarwal1990} G. S. Agarwal and R. R. Puri, Phys. Rev. A \textbf{41}, 3782 (1990).
\bibitem{Zhu2020} C. J. Zhu, L. L. Ping, Y. P. Yang, and G. S. Agarwal, Phys. Rev. Lett. \textbf{124}, 073602 (2020).
\bibitem{SM} See Supplemental Material for the derivation of Eq.~(3) from Eqs.~(1)--(2), the full linear-stability analysis, the exact two-time correlations and their numerical method, the truncated Wigner method and its validation, the proof of the $\mathbb Z_2$ theorem, numerical methods and convergence, and parameter ranges.
\bibitem{Vidal2002} G. Vidal and R. F. Werner, Phys. Rev. A \textbf{65}, 032314 (2002).
\bibitem{Plenio2005} M. B. Plenio, Phys. Rev. Lett. \textbf{95}, 090503 (2005).
\bibitem{Horodecki1998} M. Horodecki, P. Horodecki, and R. Horodecki, Phys. Rev. Lett. \textbf{80}, 5239 (1998).
\bibitem{Pikovsky2003} A. Pikovsky, M. Rosenblum, and J. Kurths, \textit{Synchronization: A Universal Concept in Nonlinear Sciences} (Cambridge University Press, 2003).
\bibitem{Minganti2018} F. Minganti, A. Biella, N. Bartolo, and C. Ciuti, Phys. Rev. A \textbf{98}, 042118 (2018).
\bibitem{Iemini2018} F. Iemini, A. Russomanno, J. Keeling, M. Schir\`o, M. Dalmonte, and R. Fazio, Phys. Rev. Lett. \textbf{121}, 035301 (2018).
\bibitem{Polkovnikov2010} A. Polkovnikov, Ann. Phys. (N.Y.) \textbf{325}, 1790 (2010).
\bibitem{Huber2021} J. Huber, P. Kirton, and P. Rabl, SciPost Phys. \textbf{10}, 045 (2021).
\bibitem{Ollivier2001} H. Ollivier and W. H. Zurek, Phys. Rev. Lett. \textbf{88}, 017901 (2001).
\bibitem{Henderson2001} L. Henderson and V. Vedral, J. Phys. A \textbf{34}, 6899 (2001).
\bibitem{Modi2012} K. Modi, A. Brodutch, H. Cable, T. Paterek, and V. Vedral, Rev. Mod. Phys. \textbf{84}, 1655 (2012).
\bibitem{MeloLuna2017} C. A. Melo-Luna, C. E. Susa, A. F. Ducuara, A. Barreiro, and J. H. Reina, Sci. Rep. \textbf{7}, 44730 (2017).
\bibitem{Muniz2020} J. A. Muniz \textit{et al.}, Nature \textbf{580}, 602 (2020).
\bibitem{Brennecke2008} F. Brennecke, S. Ritter, T. Donner, and T. Esslinger, Science \textbf{322}, 235 (2008).
\bibitem{Mottl2012} R. Mottl \textit{et al.}, Science \textbf{336}, 1570 (2012).
\bibitem{Burd2019} S. C. Burd \textit{et al.}, Science \textbf{364}, 1163 (2019).
\bibitem{Burd2021} S. C. Burd \textit{et al.}, Nat. Phys. \textbf{17}, 898 (2021).
\bibitem{Hosten2016} O. Hosten, R. Krishnakumar, N. J. Engelsen, and M. A. Kasevich, Science \textbf{352}, 1552 (2016).
\end{thebibliography}
\end{document}